\documentclass{aa}  

\usepackage{graphicx}
\usepackage{txfonts}
\usepackage{ulem}
\usepackage{graphicx}
\usepackage{xcolor}
\usepackage{lscape}
\usepackage{txfonts}
\usepackage{natbib}
\usepackage[breaklinks=true]{hyperref} 
\bibpunct{(}{)}{;}{a}{}{,}             
\hypersetup{colorlinks=true,linkcolor=magenta,citecolor=blue,filecolor=blue,urlcolor=magenta}
\newcommand{\cloudy}{\textsc{\large{Cloudy}}}
\newcommand{\hii}{H{\scriptsize II}}
\newcommand{\av}{A$_{\rm V}$}
\newcommand{\gx}{G$_{X}$}

\newcommand{\Myr}{\mbox{M$_{\odot}$ yr$^{-1}$}}
\newcommand{\Hmol}{\mbox{H$_{\rm 2}$}}

\newcommand{\vtur}{v$_{\rm tur}$}

\newcommand{\kms}{\mbox{{\rm km}\,{\rm s}$^{-1}$}}
\newcommand{\km}{\mbox{{\rm km}}} 
\newcommand{\ergcms}{erg cm$^{-2}$ s$^{-1}$}

\newcommand{\neii}{\mbox{[Ne\,{\sc ii}]$\lambda$ 12.8 $\mu$m}}

\newcommand{\neiii}{\mbox{[Ne\,{\sc iii}]$\lambda$ 15.5 $\mu$m}}

\newcommand{\oia}{\mbox{[O\,{\sc i}]$\lambda$ 63.2 $\mu$m}} 
 
\newcommand{\cii}{\mbox{[C\,{\sc ii}]$\lambda$ 157.7 $\mu$m}}

\newcommand{\niiop}{\mbox{[N\,{\sc ii}]$\lambda$ 6583 \text{\AA}}}
\newcommand{\niiopab}{\mbox{[N\,{\sc ii}]$\lambda$ 6548, 6583 $\AA$}}
\newcommand{\ha}{H${\alpha}$}
\newcommand{\oiop}{\mbox{[O\,{\sc i}]$\lambda$ 6300 $\AA$}}
\newcommand{\siiaop}{\mbox{[S\,{\sc ii}]$\lambda$ 6716 $\AA$}}
\newcommand{\siibop}{\mbox{[S\,{\sc ii}]$\lambda$ 6731 $\AA$}}
\newcommand{\oiiiop}{\mbox{[O\,{\sc iii}]$\lambda$ 5007 $\AA$}}
\newcommand{\hb}{H${\beta}$}

\definecolor{ao(english)}{rgb}{0.0, 0.5, 0.0}

\makeatletter
\renewcommand*\aa@pageof{, page \thepage{} of \pageref*{LastPage}}
\makeatother

\begin{document}

   \title{Excitation mechanisms in the intracluster filaments surrounding Brightest Cluster Galaxies}

       \author{F. L. Polles\inst{1,2}
            \and
            P. Salom\'e\inst{1}
            \and 
            P. Guillard\inst{3,4}
            \and
            B. Godard\inst{5}
            \and
            G. Pineau des For\^ets\inst{1,6}
            \and
            V. Olivares\inst{1}
            \and 
            R. S. Beckmann\inst{3}
            \and
            R. E. A. Canning\inst{7}
            \and
            F. Combes\inst{1,8}
            \and
            Y. Dubois\inst{3}
            \and
            A. C. Edge\inst{9}  
            \and
            A. C. Fabian\inst{10} 
            \and
            G. J. Ferland\inst{11} 
            \and 
            S. L. Hamer\inst{10} 
            \and
            M. D. Lehnert\inst{3}
        }

   \institute{LERMA, Observatoire de Paris, PSL research
                Universit\'e, CNRS, Sorbonne Universit\'e, 75104 Paris, France
                \and
                SOFIA Science Center, USRA, NASA Ames Research Center, M.S. N232-12 Moffett Field, CA 94035, USA\\ \email{fpolles@usra.edu}
                \and 
                Sorbonne Universit\'e, CNRS, UMR 7095, Institut d'Astrophysique de Paris, 98bis bd Arago, 75014 Paris, France
                \and
                Institut Universitaire de France, Minist\`ere de l’Education Nationale, de l’Enseignement Sup\'erieur et de la Recherche, 1 rue Descartes, 75231 Paris Cedex 05, France
                \and
                \'Ecole normale sup\'erieure, Universit\'e PSL, Sorbonne Université, CNRS, LERMA, 75005, Paris, France
                \and
                Universit\'e Paris-Saclay, CNRS, Institut d’Astrophysique Spatiale, 91405, Orsay, France
                \and
                Kavli Institute for Particle Astrophysics and Cosmology, Stanford University, 452 Lomita Mall, Stanford, CA 94305-4085, USA
                \and
                Coll\`ege de France, 11 Place Marcelin Berthelot, 75005 Paris, France
                \and
                Institute for Computational Cosmology, Department of Physics, Durham University, South Road, Durham DH1 3LE, UK
                \and
                Institute of Astronomy, University of Cambridge, Madingley Road, Cambridge CB1 0HA, UK
                \and
                Department of Physics and Astronomy, University of Kentucky, Lexington, KY 40506, USA
}

   \date{Received; accepted }
 
  \abstract
 {The excitation of the filamentary gas structures surrounding giant elliptical galaxies at the center of cool-core clusters, a.k.a BCGs
(brightest cluster galaxies), is key to our understanding of active galactic nucleus (AGN) feedback, and of the impact of environmental and local effects on star formation.}
{We investigate the contribution of the thermal radiation from the cooling flow surrounding BCGs to the excitation of the filaments. We explore the effects of small levels of extra-heating (turbulence), and of metallicity, on the optical and infrared lines.}
 {Using the \cloudy\ code, we model the photoionization and photodissociation of a slab of gas of optical depth A$_{V}\leq$30~mag at constant pressure, in order to calculate self-consistently all of the gas phases, from ionized gas to molecular gas. The ionizing source is the extreme ultraviolet (EUV) and soft X-ray radiation emitted by the cooling gas. We test these models comparing their predictions to the rich multi-wavelength observations, from optical to submillimeter, now achieved in cool core cluster.}
 {Such models of self-irradiated clouds, when reaching large enough \av, lead to a cloud structure with ionized, atomic and molecular gas phases. These models reproduce most of the multi-wavelength spectra observed in the nebulae surrounding the BCGs, not only the LINER-like optical diagnostics: \oiiiop/\hb, \niiop/\ha\ and (\siiaop+\siibop)/\ha\ but also the infrared emission lines from the atomic gas. \oiop/\ha, instead, is overestimated across the full parameter space, except for very low \av.  The modeled ro-vib \Hmol\, lines also match observations, which indicates that near and mid-IR \Hmol\, lines are mostly excited by collisions between \Hmol\, molecules and secondary electrons produced naturally inside the cloud by the interaction between the X-rays and the cold gas in the filament. However, there is still some tension between ionized and molecular line tracers (i.e. CO), which requires to optimize the cloud structure and the density of the molecular zone. The limited range of parameters over which predictions match observations allows us to constrain, in spite of degeneracies in the parameter space, the intensity of X-ray radiation bathing filaments, as well as some of their physical properties like \av\, or the level of turbulent heating rate.}
  {The reprocessing of the EUV and X-ray radiation from the plasma cooling is an important powering source of line emission from  filaments surrounding BCGs. \cloudy\ self-irradiated X-ray excitation models, coupled with a small level of turbulent heating, manage to reproduce simultaneously a large number of optical-to-infrared line ratios when all the gas phases (from ionized to molecular) are modelled self-consistently. Releasing some of the simplifications of our model, like the constant pressure, or adding the radiation fields from the AGN and stars, as well as a combination of matter- and radiation-bounded cloud distribution, should improve the predictions of line emission from the different gas phases.}
 

   \maketitle
%
\section{Introduction}
X-ray observations of galaxy clusters show that for more than one third of the clusters, the X-ray surface brightness peaks in the center. This  emission  is  due  to  the  cooling of the hot intracluster medium (ICM) with a short radiative cooling time. At the center of these ``cool-core'' clusters lies a giant elliptical galaxy, the brightest cluster galaxy (BCG). {\it Chandra} and XMM-Newton X-ray observations of the BCGs, revealed huge ICM cavities, produced by the jet of the central black hole (e.g., \citealt{mcnamara07}; \citealt{fabian15}). These cavities highlight the impact of active galactic nucleus (AGN) on their large-scale environments and suggest that the necessary heating source to prevent the overcooling in the cool-core cluster could be provided by the AGN.  
Observations at different wavelengths of these regions, show that the cavities are often surrounded by multi-wavelength line-emitting filamentary structures (e.g., \citealt{heckman89}; \citealt{conselice01}; \citealt{lim12}; \citealt{mittal12}; \citealt{rose20}), illustrating the multi-phase nature of these streams. Studies of the correlation between H$\alpha$ and CO(1-0) line emission (e.g. \citealt{salome11, tremblay18, olivares19, russell19}) showed that these two tracers are co-spatial and co-moving with most of the mass of the filaments lying in the molecular phase. These filaments may have formed from the cooling of the hot gas in very local regions around the AGN-cavities, when the gas is locally thermally unstable (e.g. \citealt{gaspari12}; \citealt{beckmann19}; \citealt{qiu20}).

In the last decades several studies have investigated the powering source of the multi-wavelength spectrum of these filamentary nebulae (e.g.~\citealt{heckman89}; \citealt{voit94}; \citealt{ferland94, ferland02, ferland08}; \citealt{bayet11}; \citealt{canning16}), proposing and exploring different sources of photoionization and heating: (1) the central AGN, (2) X-rays from ICM, (3) heat conduction from the ICM to the cold filament, (4) shocks and turbulent mixing layers, (5) collisional heating by cosmic rays, (6) hybrid models (several energy sources). Some of these sources have already been excluded: photoionization by the AGN is not powerful enough to produced the observed H$\alpha$+\mbox{[N\,{\sc ii}]} line emission (e.g. \citealt{heckman89}; \citealt{conselice01}) and disagree with the lack of strong radial gradients profile in the optical emission line ratios (e.g. \citealt{heckman89}). The fairly constant H$\alpha$/H$_2$ \citep{lim12} and  H$\alpha$/CO (e.g. \citealt{olivares19}) over the entire nebula, also suggests that the excitation process is local. A local ionization mechanism could be the photoionization by young massive stars, but this scenario is not favored either, since the observations show strong \niiop/\ha\ ratio and weak \oiiiop/\hb\ ratio, indicating that the optical spectra of these nebulae are more similar to low ionization nuclear emission region (LINER) spectra than \hii\, spectra (e.g. \citealt{heckman87}; \citealt{crawford92}). \cite{mcdonald12} claimed that a model with a mix of heating by shocks (100 - 400 km s$^{-1}$; \citealt{allen08}) and photoionization by stars (\citealt{kewley01}) can reproduce the optical line ratios \niiop/\ha, \oiop/\ha, (\siiaop+\siibop)/\ha, and \oiiiop/\hb, of a sample of nine BCGs. 
But this study was limited to optical lines only. In the near infrared, \cite{jaffe97} investigated the line ratio H$_{2}$ 1-0 S(1)/H$\alpha$. They measured a ratio of $\sim$0.1, which is larger than the observed ratio in Galactic \hii\, regions ($\sim$0.01) and excludes fast shocks (>50 \kms) as possible excitation mechanism. The possible source of excitation of the ro-vibrationally excited H$_{2}$ lines in the filaments have been discussed also by \cite{wilman02} and \cite{lim12}.
 
Focusing on one filament around NGC\,1275, the so-called Horseshoe-region, \cite{ferland08,ferland09} extended the observational constraints from optical to infrared, and proposed as the main heating mechanism, the collisions of the cold gas in the filaments with ionizing particles. The authors showed that this model is consistent with the filaments being made of cloudlets of varying densities and excited by cosmic-rays. One strong motivation and success of this model was to reproduce the surprisingly strong molecular hydrogen lines detected by {\it Spitzer} in Perseus and Centaurus filaments (\citealt{johnstone07}), together with the ratios of the near infrared (NIR) H$_{2}$ line emission, as well as the lack of \oiiiop\, emission line noticed in most of the filaments of Perseus by \cite{hatch06}. 
\cite{fabian11} showed that the surrounding hot ICM is a possible source of these energetic particles. \cite{bayet11} used the same kind of model (energetic particles with/without extra-heating) to predict molecular emission lines like  CN (2--1), HCO$^{+}$ (3--2) and C$_{2}$H (3--2). With the advent of {\it Herschel Space Observatory}, the detection of the far-infrared (FIR) \cii\, and \oia\, line emission, was made possible in BCGs and even mapped in the filaments of the Perseus and Centaurus clusters. \cite{mittal12} showed that the observed ratio \oia/\cii\ was barely reproduced by the energetic particles model only. In order to reproduce this ratio with such model, it is necessary to add an extra heating source such as turbulent heating, with typical velocity dispersion of 2-10 \kms as shown in \cite{canning16}. This model succeed on reproducing the observations, however, it is a result of a weighted sum of a power-law distribution of cloud densities for which the \cloudy\ computation was stopped at the illuminated face of the cloud (first zone), i.e. only emission from the skin of the clouds was combined. In \cite{canning16} the authors investigated the behavior of the predicted line emissivity throughout a cloud at fixed density, but not for an integral of clouds of many densities.

In this paper we re-investigate the effect of the excitation due to thermal radiation as expected from the cooling of the hot plasma and we explore the outcomes due to photoionization at different depth of the cloud, namely the visual extinction (\av). The photoionization by the X-ray photons emitted by the condensing and cooling gas, the so-called cooling flow, was already considered (i.e. \citealt{voit90};~\citealt{ferland94};~\citealt{donahue91} and references therein). 
The cool-core clusters have measured mass deposition rates of  10 - 100~\Myr, or smaller when considering regions inside the filaments. Even if large amount of energy are expected via this cooling gas, \cite{crawford92} argued against this process to power the nebula. From unresolved X-ray data, they showed that the photoionization by the surrounding X-ray gas was not powerful enough to excite the nebula of the core of the Perseus cluster. Recent deep and high resolution observations of X-ray filaments by the satellite {\it Chandra} (\citealt{walker15}) show an average intensity of twelve filaments in Perseus of 4$\times$10$^{-16}$ \ergcms arcsec$^{-2}$, made this mechanism interesting to explore again. In the present study, we do not add any other source of excitation in order to measure and discuss the effects of reprocessing the X-ray cooling energy from the ionized to the molecular phase. We are aware that this model is over-simplified to be able to reproduce all of the gas phases simultaneously. We emphasize that we explore the physical properties of the cloud and the predicted line emission at different \av. We also discuss the impact of adding some turbulent heating and of varying the metallicity (Z) on the physical and chemical properties of the cloud. We compare model predictions with observations of line emission from the filaments, combining the state-of-the-art models with the recent observations at different wavelengths, from the optical to the infrared, extending the range of constraints and predictions for a single model.

The model setup is described in section~\ref{sec:model}. The effects on varying the different free parameters are analyzed in sections~\ref{sec:physicsmodels} and ~\ref{sec:cum_emission}. The predicted optical-to-submillimeter line emission from our grid of models are compared with the observations provided by the literature in section~\ref{sec:results}. Our results are discussed in section~\ref{sec:discussion} and the conclusions are summarized in section~\ref{sec:conclusions}.

\section{Modeling}
In this section we present the setup of the \cloudy\, models as well as the evolution of the chemical and physical structure of the cloud with the free parameters of the models. 

\subsection{Setup of the models}\label{sec:model}
We use the photoionization and photodissociation code \cloudy\ v17.01 (\citealt{ferland17}), which allows to calculate self-consistently the thermal and chemical structure of a plane-parallel layer of a gas and dust. Previous studies focused the analysis on the neutral and molecular phases, however, the initial conditions of each phase are a consequence of the processes that take place on the previous phase. Moreover, some line emission, such as \cii, can arise from both ionized and neutral phase. Hence, computations that do not take into account multiple gas phases could bring incorrect interpretation to the phase properties. In our analysis, each model is performed at a constant total pressure\footnote{In \cloudy, the total pressure includes ram, magnetic, turbulent, particle, and radiation pressure. More information can be found in \cloudy's manual (\href{https://www.nublado.org/}{www.nublado.org}).} of 10$^{6.5}$ K\,cm$^{-3}$, which is the average pressure of the hot gas surrounding the filaments of NGC\,1275, as deduced from X-ray observations (\citealt{sanders07}). Similar pressure values have been estimated from the electron density calculated with the line ratio [S\,{\sc ii}]$\lambda\lambda$ 6716,6731 $\AA$ (e.g. \citealt{heckman89}). All models are stopped at the visual extinction of \av\,= 30 mag. Other input parameters for the models are: the shape and intensity of the incident radiation field, the chemical composition of dust and gas, the metallicity (Z) and the turbulent velocity dispersion (v$_{\rm tur}$). \av\,=\,30 mag correspond to slightly different physical size, based on the initial conditions of the model. To give an order of magnitude, for a model of \gx\,=10, v$_{\rm tur}$=10 \kms\, and Z\,=\,Z$_{\odot}$ the size of the full cloud is $\sim$48 pc.
The model parameters are described in detail below and a summary is provided in Table~\ref{tab:modelSummary}. 

\begin{table}[htbp!]
        \caption{Model summary.} 
        {\small \begin{tabular}{p{3.2cm} p{5.2cm}}
       		\hline
       		\hline
        		 \noalign{\smallskip}
       			Fixed parameters\\
		 \noalign{\smallskip}
		 \hline
		\noalign{\smallskip}
			Geometry & 1D plane-parallel\\
			\noalign{\smallskip}
			Radiation field & intergalactic background continuum\\  
			& HM05$^{a}$ at redshift 0\\
			\noalign{\smallskip}
				& Table SED ``cool.sed''$^{(d)}$ (X-ray emission), intensity (\gx$^{c}$) set to values below\\
			\noalign{\smallskip}
            & ISRF$^{b}$ scaled to G$_{0}^{c}$=10$^{-5}$\\
			\noalign{\smallskip}
			& CMB\\
			\noalign{\smallskip}
			& cosmic ray background: H$^{0}$ ionization rate of 2$\times$10$^{-16}$ s$^{-1}$\\
			\noalign{\smallskip}
			Density law & constant pressure set to 10$^{6.5}$ K\,cm$^{-3}$\\
			\noalign{\smallskip}
		\hline
       		\hline
        		 \noalign{\smallskip}
       			Varied parameters & values\\
		 \noalign{\smallskip}
		 \hline
		\noalign{\smallskip}
		X-ray radiation field intensity (\gx) & [10$^{-2}$, 10$^{-1.8}$, 10$^{-1.6}$, 10$^{-1.4}$, 10$^{-1.2}$, 10$^{-1}$, 10$^{-0.8}$, 10$^{-0.6}$, 10$^{-0.4}$, 10$^{-0.2}$, 1, 10, 10$^{1.2}$, 10$^{1.4}$, 10$^{1.6}$ , 10$^{1.8}$, 10$^{2}$, 10$^{2.2}$, 10$^{2.4}$, 10$^{2.6}$, 10$^{2.8}$, 10$^{3}$]\\
		Metal abundances (Z) & [0.3, 0.35, 0.4, 0.45, 0.5, 0.55, 0.6, 0.65, 0.7, 0.75, 0.8, 0.85, 0.9, 0.95, 1] Z$_{\odot}$\\
		Turbulent velocity (v$_{\rm tur}$) & [0, 2, 10, 30, 50, 100] \kms\\
	 \noalign{\smallskip} 
	 \hline
	 \hline
\end{tabular}} 
\\
{\small{\bf Notes.} $^{(a)}$ Haardt $\&$ Madau radiation field available in \cloudy. $^{(b)}$Interstellar radiation field calculated by Meudon PDR code (\citealt{lepetit06}) using the radiation field from \cite{mathis83}. $^{(c)}$\gx\,=\,1 (respectively G$_{0}$\,=\,1) corresponds to an
integrated intensity between 0.2 and 2 keV (respectively  6 - 13.6 eV) of 1.6$\times$10$^{-3}$\ergcms. $^{(d)}$SED shape available in \cloudy\, to reproduce an X-ray emission from cooling flow gas.} 
 \label{tab:modelSummary}
\end{table} 

\subsubsection{Input radiation field: shape}\label{sec:SEDshape}
One of the key parameters of the present modeling is the shape of the input radiation field. Energetic particles or hard X-ray fail to reproduce the observed \oiiiop/\hb\, ratio, because they produce too low highly ionized oxygen \oiiiop\, with respect to the other ionized species. To solve this problem, extreme ultraviolet (EUV) as well as soft X-ray photons are necessary since they manage to create and excite the different ions. The slope
of the input field has thus a direct impact on the ionized line ratios. Such photons can be emitted by the gas in cooling
flows (self-irradiation). This model has been described in \cite{johnstone92}. The total emission due to the gas
that cools constantly from the temperature of the hot surrounding medium, is the result of the sum of the
gas emission at each temperature normalized by the mass deposition rate. In our models, the soft X-ray/EUV band of
the input radiation field is thus a power-law resulting by the co-added series of Raymond-Smith collisional-equilibrium
continua (\citealt{raymond77}). This continuum shape is available in \cloudy\footnote{This shape can be used
as input SED with the command {\it Table
SED ``cool.sed''}.} as an input spectral energy distribution (SED) shape and it has been already used in \cite{ferland94,ferland02} to model cooling flow environments. Fig.~\ref{fig:SEDout} shows that almost all of this soft X-ray, EUV radiation is self-absorbed and reprocessed at larger wavelength when it encounters a cloud
with a large enough depth. As explained in section ~\ref{sec:intensity}, we fine-tune the input radiation intensity and the \av\, to reproduce the X-ray surface brightness observed in the filaments of Perseus. Our simplified model do not explore the spatial distribution or the size distribution of the clouds. We note that one strong hypothesis of this model is that a large fraction of the hot cooling gas radiation is absorbed and reprocessed by the slab of atomic and molecular gas. Our models explore a range of cloud optical depths, which allows us to both match the observed X-ray fluxes and multi-wavelength line emission. We leave for a further work a more complex model of a fog of atomic and molecular clouds spread in a bath of hot cooling gas.

We also added an intergalactic background continuum, taken to be the 2005 version of the \cite{haardt96} background at redshift 0, with both starburst and quasar continua. The input SED also includes the cosmic microwave background (CMB) and the infrared dust emission of the standard interstellar radiation field. 

\subsubsection{Input radiation field: intensity}\label{sec:intensity}
The intensity of the X-ray/EUV emission is
varied with the parameter \gx. By definition \gx\,=\,1 corresponds to an integrated intensity between 0.2 - 2 keV of
4$\pi\nu$F$\nu$ = 1.6$\times$10$^{-3}$ erg cm$^{-2}$s$^{-1}$. To compare the input and output values of the radiation field in $\cloudy$ with the observed surface brightness, one has to divide this values by 4$\pi$ in order to get values in erg cm$^{-2}$ s$^{-1}$ sr$^{-1}$. For instance \gx\,=\,1 corresponds to an input field of 1.3$\times$10$^{-4}$ erg cm$^{-2}$s$^{-1}$ sr$^{-1}$, that is 3$\times$10$^{-15}$ erg cm$^{-2}$s$^{-1}$ arcsec$^{-2}$.

We compared the $\cloudy$ outward radiation field in the range 0.6-2.0 keV with the X-ray surface brightness observed in the same band by \cite{walker15} in the filaments of the Perseus cluster. The authors estimated that the average surface brightness in this band of the filaments is 4.0$\times$10$^{-16}$ ergs cm$^{-2}$ s$^{-1}$ arcsec$^{-2}$. This is the red dot at 2.1$\times$10$^{-4}$ erg cm$^{-2}$s$^{-1}$ on Fig.~\ref{fig:SEDout}. To give orders of magnitudes, \cite{sanders05} estimated the surface brightness of the hard X-ray emission in the band 2-10 keV and found that it decreases with the radius, from $\sim$5$\times$10$^{-15}$ ergs cm$^{-2}$ s$^{-1}$ arcsec$^{-2}$ at radius 20 arcsec (magenta dot on Fig.~\ref{fig:SEDout}) to 10$^{-16}$ ergs cm$^{-2}$ s$^{-1}$ arcsec$^{-2}$ at radius 200 arcsec.

In our grid of models, \gx\,varies between 10$^{-2}$ and 10$^{3}$, with step of 0.2 dex. Note that the range of \gx\,that is explored here is higher than in \cite{ferland94,ferland02} where self-irradiated clouds had also been modeled. The values used at that time were scaled on unresolved X-ray observations. The more recent X-ray observations in Perseus are by \cite{sanders07}; \cite{walker15} showed that the X-ray radiation is not homogeneously spread over the entire 100
kpc region and soft X-ray emission arising from the filaments surrounding NGC\,1275 have surface brightness values
reaching up to a few 10$^{-16}$ ergs cm$^{-2}$ s$^{-1}$ arcsec$^{-2}$ as mentioned above. The comparison between the input SED used by \cite{ferland94} and the input SED of one of our models is shown in Figure~\ref{fig:ferlandSED}. Figure~\ref{fig:SEDout} shows that low \gx\ are too small compared to the X-ray observations in Perseus. We nevertheless include grids of model with such low \gx\ in order to show the effect of turning-on \gx\ radiation.

\subsubsection{Outward radiation field: cloud optical depth effects}\label{sec:av}
\begin{figure*}[htbp!]
        \includegraphics[trim={0cm 1cm 0cm 0cm}, clip, width=\hsize]{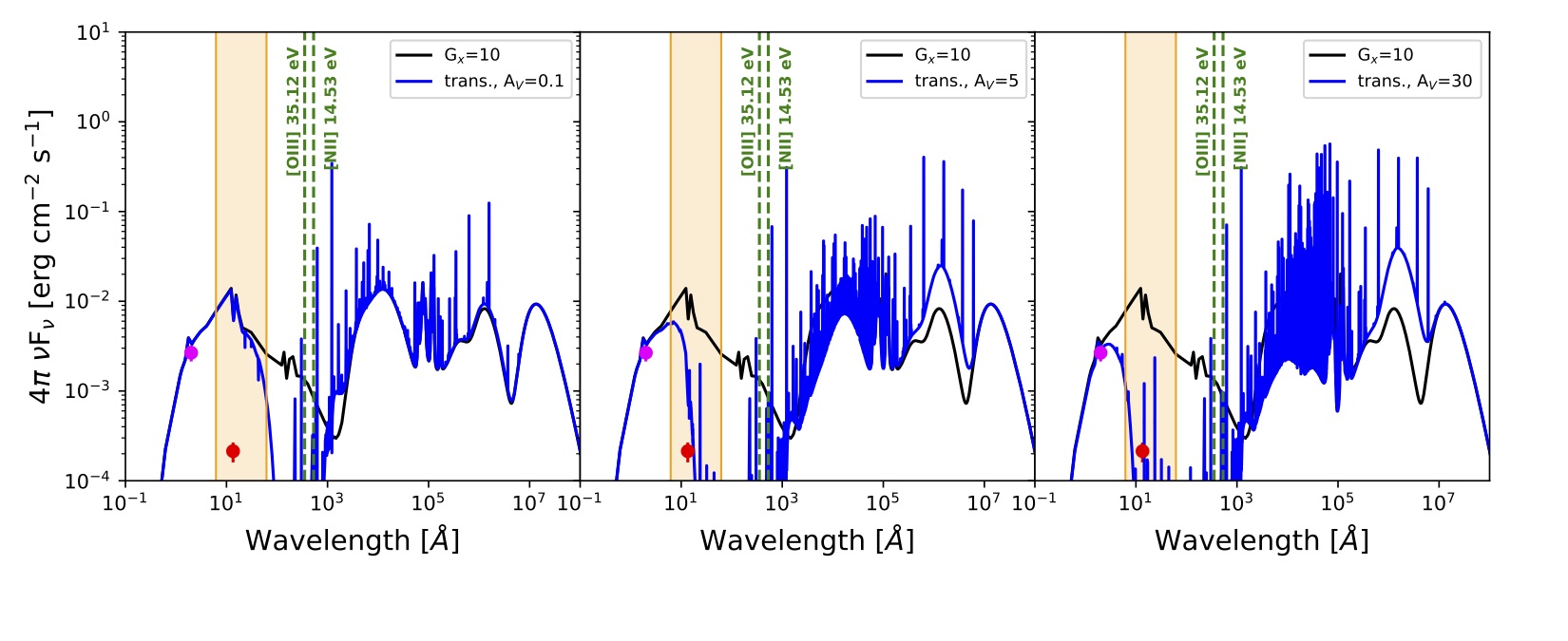}
        \caption{SED of the input radiation field for \gx\,=\,10 (black line) and the total outward radiation field (continuum and line emission) stopping the model at different \av (blue line): 0.1, 5 and 30 mag. The two vertical green lines show the energy necessary to create the ion O$^{++}$ and N$^{+}$, two of the ions whose emission are used to constrain the modeling. The dots represent the average observed integrated flux for the filaments surrounded NGC\,1275 of 4.0$\times$10$^{-16}$ \ergcms arcsec$^{-2}$ in the 0.6-2.0 keV band (red dot, \citealt{walker15}), and in the 2-10 keV band (magenta dot) of 5$\times$10$^{-15}$ \ergcms arcsec$^{-2}$ (\citealt{sanders05}). The orange background highlights the range over which the integrated intensity is calculated to scale \gx.}
    \label{fig:SEDout}
\end{figure*}

The new aspect of our modeling is that we self-consistently compute the chemical, thermodynamical structure and emissivity of the entire cloud as a function of \av~$\leq30$~mag. Figure~\ref{fig:SEDout} shows how the outward, transmitted spectrum changes with the total\footnote{Value at which the model calculation is stopped.} \av\, of the cloud. With the increase of \av, i.e. of the hydrogen column density, the cloud absorbs more continuum in the energy range between X-ray to near-infrared (NIR), all of the hydrogen-ionizing radiation is absorbed and the dust emission in FIR increases. Stopping the calculation at \av=30 mag implies to go throughout different gas phases. 

\subsubsection{Elemental abundances}
The chemical setup is based on the previous studies of \cite{ferland09} and \cite{canning16}. The gas elemental abundances and the chemical composition of dust are set to those derived in the Orion Nebula by \citet{baldwin91}, \citet{rubin91}, \citet{osterbrock92}, and \citet{rubin93} (see Tables~\ref{tab:abundances} and \ref{tab:grains}). The grain size distribution is set to the standard distribution in the Milky Way presented in \cite{weingartner01}, which consists of a mixture of graphites and silicates and is known to reproduce the standard Galactic extinction curve with $R_{\rm V}$ = 3.1. The polycyclic aromatic hydrocarbon (PAH) size distribution is given by a power law 
described in \cite {abel08}. 

While most parameters are calibrated with Galactic values, it is important to note that the elemental composition and dust properties in the filaments could be different. Indeed, the average metallicity of the ICM is $\sim$0.3\,$Z_{\odot}$. However, and as shown by \citet{sanders07} (Figure 10), the metallicity of the filaments varies with radius, reaching a maximum at a distance of 40 kpc. In their study the metallicity covers a range between 0.5 to 0.7 $Z_{\odot}$. To investigate in our models the effect of the metallicity on the predicted line emission, which will be compared with the line emission arising from the filaments, we explore here values between 0.3 to 1 $Z_{\odot}$ with a step of 0.05.
 
\subsubsection{Turbulence}
The input of mechanical energy in filaments surrounding BCGs originates from various processes such as the cascade of kinetic energy injected at large scales by AGN jets and inflated bubbles (e.g. \citealt{revaz08}; \citealt{beckmann19}), the turbulent mixing between the hot and cold phases (\citealt{begelman90}; \citealt{banerjee14}; \citealt{hillel20}), and the collisions between intertwined filaments. All these processes generate small scale turbulence which not only induces pressure fluctuations but may also lead to the propagation of low velocity molecular shocks that could be responsible for bright \Hmol\ emission of filaments \citep{johnstone07, guillard12}. In \cloudy, turbulence is modelled through a velocity dispersion parameter \vtur\, which acts as a pressure support, an homogeneous heating rate, and a contribution to the line broadening for the radiative transfer. In the collisional ionization models presented in \citet{canning16}, the authors considered turbulent velocity dispersion of $2 - 10$~\kms, which is enough to decrease the optical depths of the lines and improve the cooling efficiency. 

The lines originating from the warm gaseous nebula surrounding NGC\,1275 have typical full width at half-maximum (FWHM) ranging between 50 and 200~\kms\ (\citealt{lim12}). In comparison, a FWHM of $\sim$30~\kms\ was measured from CO(1-0) in the same region \citep{salome08}. More recently, FWHMs of 100 to 140~\kms\ were obtained from several emission lines of CO for a sample of various BCGs \citep{olivares19} and FWHMs of 3 to 150~\kms\ from several molecular absorption lines seen against the bright  radio core of Hydra-A \cite{rose19, rose20}.Whether these line widths originate from the internal velocity dispersion of cold filaments or result from relative motions between filaments is an unsolved issue. In the latter case, the velocity dispersion of the cold components would be intrinsically smaller. To explore all possible scenarios, we extend the range of turbulent velocity dispersion studied by \citet{canning16} and adopt values of v$_{\rm tur}$ between 0 and 100~\kms.

\subsection{Thermal and ionization profiles}\label{sec:physicsmodels}

\begin{figure*}[htbp!]
\includegraphics[trim={1cm 0cm 1cm 0cm}, clip, width=\hsize]{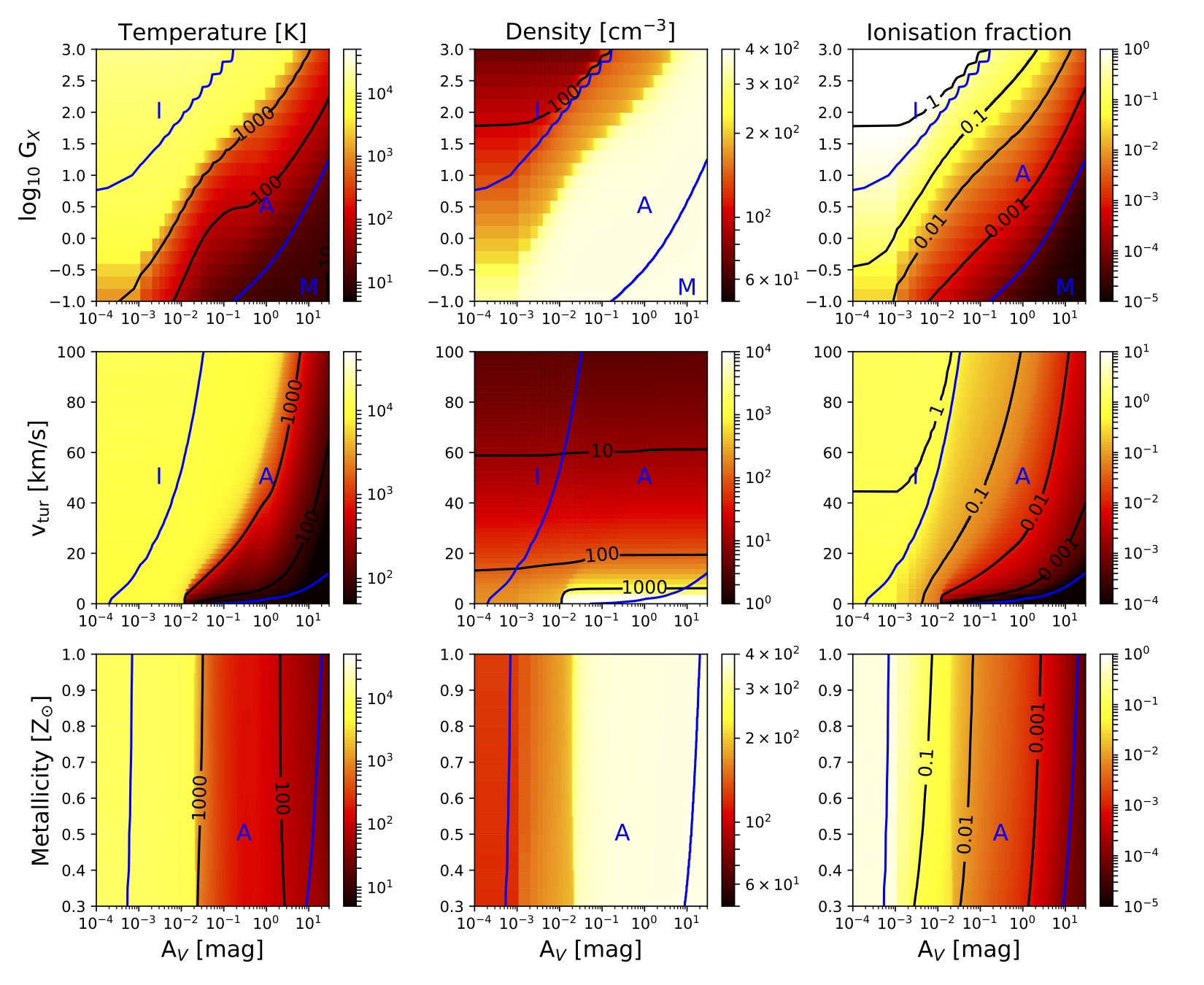}
\caption{Temperature (left column), hydrogen density (middle column), and ionization fraction $n_{e^{-}}$/$n_{\rm H}$ (right column) of the gas predicted by \cloudy\ as functions of the visual extinction \av\ and the X-ray radiation field scaling factor \gx\ (top row), the turbulent velocity dispersion \vtur\, (middle row), and the metallicity $Z$ (bottom row). Non varying parameters are set to their standard values. In particular the total pressure of the gas is set to 10$^{6.5}$ K cm$^{-3}$, G$_{X}$=10 (for middle and bottom rows), \vtur\,=10 \kms\ (for top and bottom rows), and $Z=Z_{\odot}$ (for top and middle rows). The black curves correspond to isocontours. The transition from ionized to atomic phase and that of atomic to molecular defined by $n_{e^{-}}$/$n_{\rm H^{0}}$=1 and to $n_{\rm H^{0}}$/$n_{\rm H_{2}}$=1, respectively, are indicated with a blue curve. The blue I, A and M identify the different gas phases: ionized, atomic and molecular, respectively.}
    \label{fig:propxr_int}
\end{figure*}

\begin{figure*}[htbp!]
        \includegraphics[trim={2cm 6.5cm 0cm 7cm}, clip, width=\hsize]{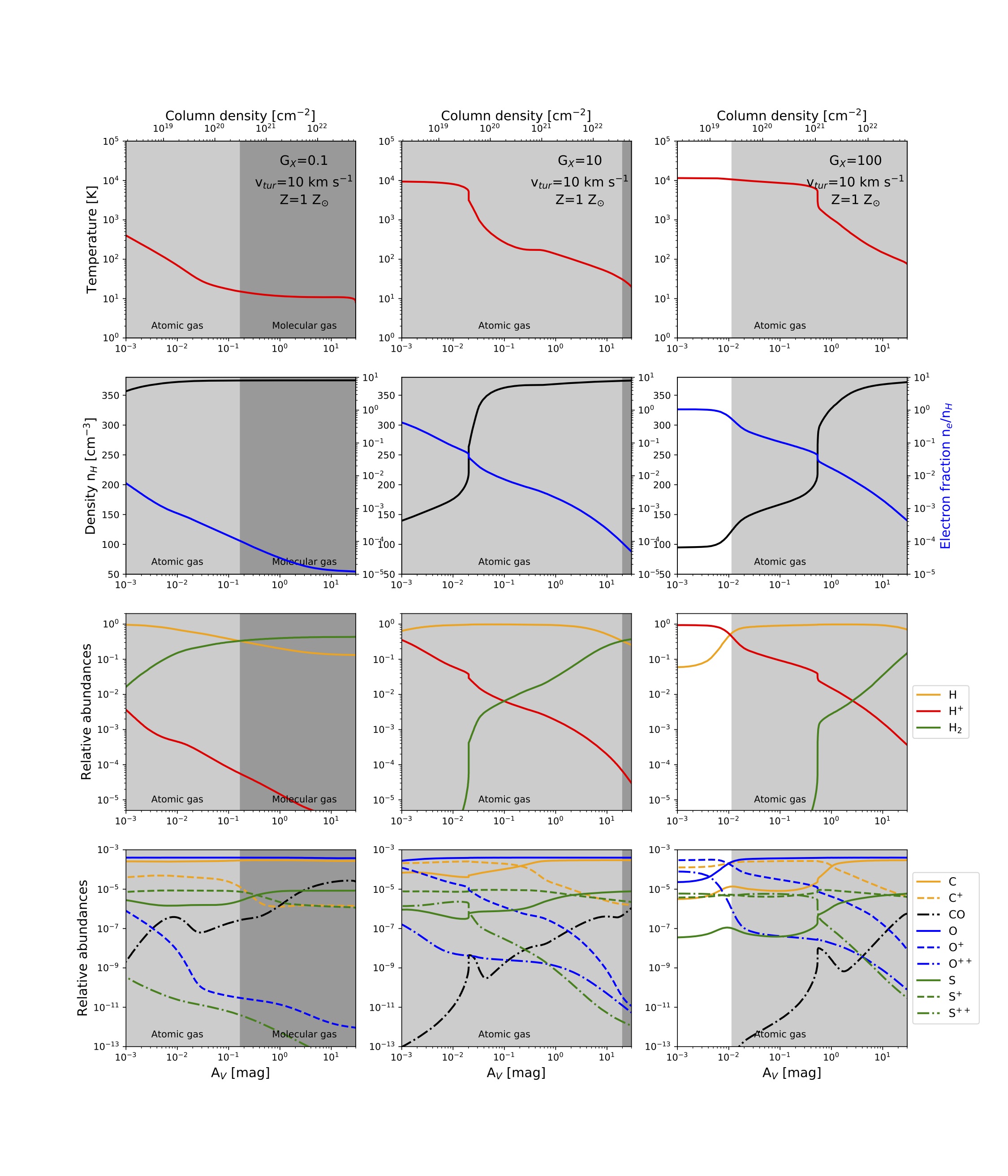}
        \caption{Effects of the radiation field intensity. Physical properties as a function of \av\ for models in thermal equilibrium and constant pressure fixed at 10$^{6.5}$ K cm$^{-3}$, with solar metallicity,  turbulent velocity 10 km s$^{-1}$ and different values of X-ray radiation field intensity: \gx\,= 10$^{-1}$({\it left column}), \gx\,= 10$^{1}$({\it central column}) and \gx\,= 10$^{2}$ ({\it right column}). {\it Top row}: Gas temperatures. {\it Second row}: total hydrogen density in black and ionization fraction (n$_{\rm e}$/n$_{\rm H}$) in blue. {\it Third row}: relative abundances of hydrogen. {\it Bottom row}: relative abundances of carbon (yellow), oxygen (blue), sulfur (green) and CO (black). The white background indicates the ionized phase, the gray background the atomic phase and the dark gray the molecular phase. The transition from ionized to atomic phase and that of atomic to molecular correspond to density ratio of e$^{-}$/H$^{0}$ = 1 and of H$^{0}$/H$_{2}$=1, respectively.}
    \label{fig:exproplx1}
\end{figure*}

The main parameters of the models are the intensity of the input X-ray radiation field, the turbulence strength, and the metallicity. Figure~\ref{fig:propxr_int} summarizes how these parameters influence the thermal and ionization structures of a cloud of constant total pressure $P=10^{6.5}$ K cm$^{-3}$. To facilitate the interpretation, we also display in Figs. \ref{fig:exproplx1}, \ref{fig:exproplx1tur}, and \ref{fig:exproplx1met}, the main heating and cooling processes, and the chemical profiles associated to seven models representative of the entire parameter domain.

Because X-ray photons dominate both the heating and the ionization rates of the gas, the structures shown in Fig. \ref{fig:propxr_int} are similar to the standard profiles predicted in photo-dominated regions (PDRs). As a function of the visual extinction $A_{\rm V}$, the cloud undergoes a transition from a hot ($T \sim 10^4$ K), diffuse, and partially or fully ionized phase to a cold ($T \sim 10-100$ K), dense, and partially or fully molecular medium. High temperatures at the border of the cloud result from an equilibrium between the heating induced by photoelectric effect on gas particles and the cooling induced by electronic transitions. In contrast, low temperatures inside the cloud result from a balance between the heating due to photoelectric effect on grains and the cooling induced by fine structure lines and molecular lines. The fact that the photoelectric effect is still efficient at large $A_{\rm V}$ is a known property of X-ray dominated regions (XDRs): because hard X-rays have small interaction cross-sections, they penetrate deep in the cloud where they locally induce the production of UV photons that participate to the ionization of dust and gas, and photodissociation.

The transition between the two thermochemical states is driven by the absorption of the X-ray radiation field. Again, because of the low interaction cross-sections of high energy photons, this transition requires a larger total column density than that required in clouds illuminated only by UV radiation field \citep{meijerink06, meijerink07}. Moreover, because X-rays are mostly absorbed by atoms, self-shielding effects are important. It follows that the cloud structure do not only depend on $A_{\rm V}$ but also on the  density profile from the border of the cloud: the smaller the density, the widest the transition. If $G_{\rm X}$ increases, both the ionization and molecular fronts necessarily occur at larger extinction. This is due to the increase of the input X-ray flux at high energy, the decrease of density at the border to ensure a constant total pressure, and the local production of UV photons induced by the thermalization of photoelectrons.

In this framework, the impact of the turbulent velocity dispersion is straightforward. Because the total pressure is assumed to be constant, the mass density is
\begin{equation}
\label{eq:cloudy_rho}
\rho = \frac{P}{\frac{{\rm v}_{\rm tur}^2}{2} + \frac{k T}{\mu m_{\rm H}}},
\end{equation}
where $P$ is the total pressure, $\mu$ is the mean molecular mass of the gas, and  $m_{\rm H}$ the mass of hydrogen atoms.  Therefore, the primary effect of the velocity dispersion in \cloudy\ is to globally reduce the density of the gas over the entire cloud. Equation \ref{eq:cloudy_rho} also reveals a threshold effect: as \vtur\, increases, the turbulence pressure (P$_{\rm tur}$) increases and the thermal gas pressure (P$_{\rm th}$) decreases. When the P$_{\rm tur}$ becomes higher than the P$_{\rm th}$, the cloud  switches from a medium at constant thermal pressure with sharp profiles to a medium at constant density with smoother profiles. Such transition occurs for P$_{\rm tur}$ > P$_{\rm th}$, thus
\begin{equation}
{\rm v}_{\rm tur} \geqslant {\rm v}_{\rm tur}^{\rm lim} = 1.6 \,\, \kms\ \left( \frac{T}{100\,\,{\rm K}} \right)^{0.5} \geqslant 16 \,\, \kms,
\end{equation}
in agreement with the middle panel of Fig. \ref{fig:propxr_int}. Finally, since $\rho$ decreases when \vtur\, increases, both the ionization and molecular fronts necessarily shift towards larger $A_{\rm V}$.

As shown in Figures~\ref{fig:propxr_int} and \ref{fig:exproplx1met}, the metallicity appears to have a very limited impact on the thermochemical properties of the cloud. This result is slightly misleading because the metallicity is explored over a narrow range of values. When the metallicity decreases, the abundances of dust and heavy elements decrease by the same factor. A given $A_{\rm V}$ therefore corresponds to higher hydrogen column density, hence the ionization front occurs correspondingly sooner.

\subsection{Cumulative emission}\label{sec:cum_emission}

The physics of a cloud illuminated by high energy photons is driven by the reprocessing of the input radiation field into continuum and line emission. A PDR (illuminated by UV photons) and an XDR (illuminated X-ray photons) differ by the nature of this reprocessing and the amount of matter it requires. In classical PDRs, the input energy flux is reprocessed over typical visual extinction $A_{\rm V}\sim 1$, mostly through continuum dust emission; only a few percent is converted into atomic and molecular lines. In contrast, and because the impinging radiation field consists of photons of higher energies (X-ray; see Fig. \ref{fig:SEDout}), the reprocessing of the input radiative flux in XDRs not only requires large total column densities ($A_{\rm V}$ typically larger than 30 magnitudes), but also occurs through efficient atomic and molecular line emissions.

\begin{figure*}[htbp!]
        \includegraphics[trim={2cm 5cm 0cm 5.5cm}, clip, width=\hsize]{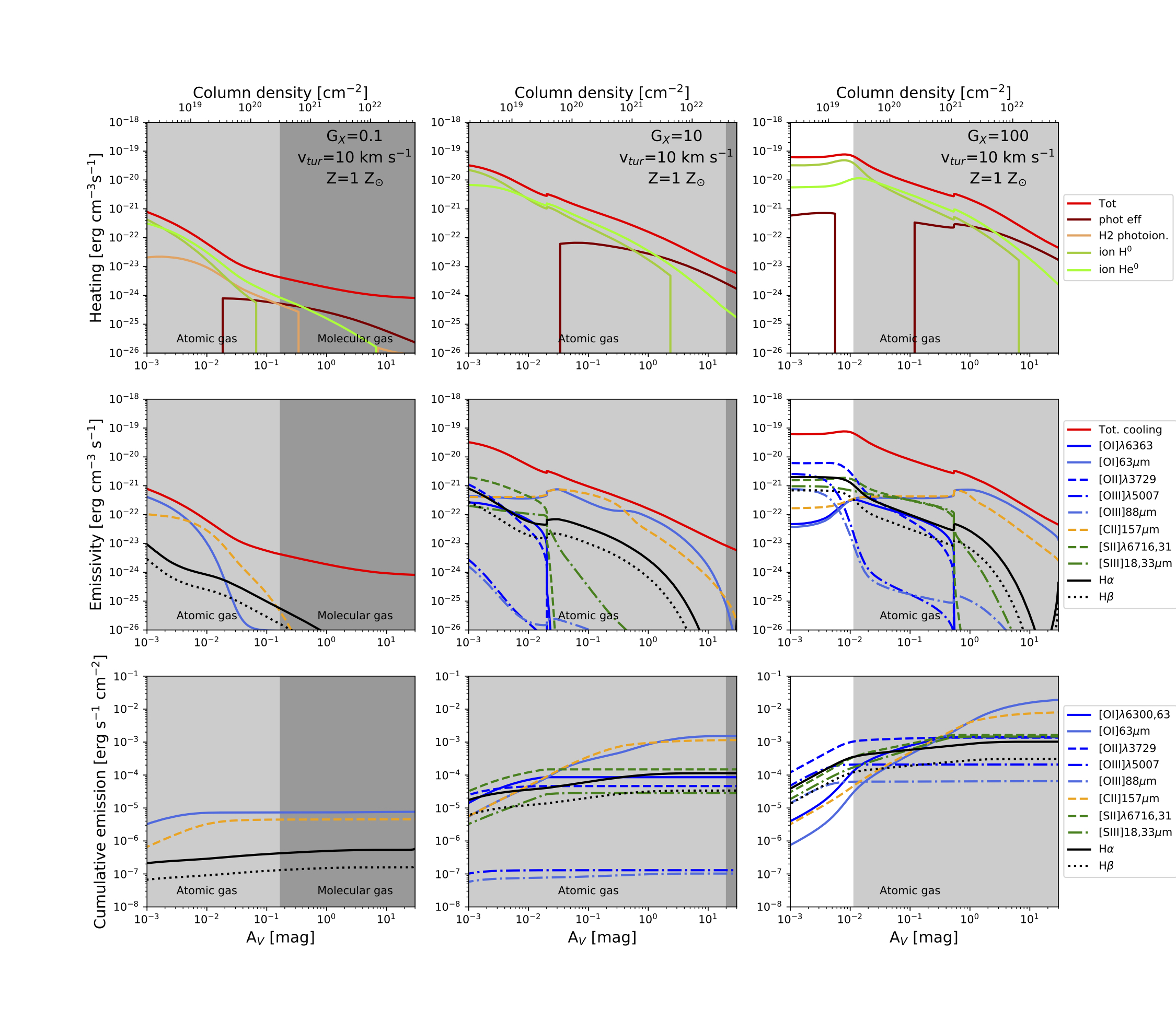}
        \caption{Effects of the radiation field intensity on heating mechanisms and line emission. Heating and cooling mechanisms as a function of \av\ for the same models as those in Fig.~\ref{fig:exproplx1}. {\it Top row}: heating mechanisms. \cloudy\ gives in output the first five main heating mechanisms. If one of the heating mechanisms is not one of them for some values of \av, the mechanism is not shown in the figures. This is the reason of some vertical straight lines in the panels. {\it Second row}: emissivity of carbon, oxygen and sulphur lines, \ha\, and \hb. {\it Bottom row}: cumulative emission of carbon, oxygen and sulphur lines, \ha\, and \hb. The cumulative emission are theoretical values, and some of this values are too faint to be detected by the instruments. The instrumental limitations give access only to the brightest values, thus to those emission arising from the denser part of the cloud. It is important to remember that comparing the observations with these predicted line emission we constrain the properties of the denser phase of the cloud.}
    \label{fig:exproplx2}
\end{figure*}

Following the definitions of \citet{meijerink05}, the X-ray energy flux impinging the cloud is
\begin{equation}
F_{\rm X} = 1.6 \times 10^{-3} G_{\rm X} \,\, {\rm erg}\,\,{\rm cm}^{-2}\,\,{\rm s}^{-1}.
\end{equation}
The resulting absorption and emission processes are shown in Figs. \ref{fig:exproplx2}, \ref{fig:exproplx2tur}, and \ref{fig:exproplx2met} where we display the local cooling and the cumulative emissions of fine-structure, metastable, and electronic lines of several atoms and ions for different values of \gx, \vtur\, and $Z$.

The excitation of \ha\ and \hb\ occurs through collisions with high energy secondary electrons and through recombination of H$^+$. As a result, the integrated fluxes of these lines are simply proportional to the input radiation field,
\begin{equation}
F({\rm H}\alpha) \sim 10^{-5} G_{\rm X} \,\, {\rm erg}\,\,{\rm cm}^{-2}\,\,{\rm s}^{-1}
\end{equation}
and
\begin{equation}
F({\rm H}\beta) \sim 3 \times 10^{-6} G_{\rm X} \,\, {\rm erg}\,\,{\rm cm}^{-2}\,\,{\rm s}^{-1},
\end{equation}
regardless of the metallicity or the density profile set by \vtur\, (Figs. \ref{fig:exproplx2tur} \& \ref{fig:exproplx2met}). Because of these simple relations, \ha\ and \hb\ are valuable proxys of the input radiation field and can thus be used as normalization factors for other atomic and ionized lines. The results of \cloudy\ on the optical line ratios are presented in Sec.~\ref{sec:optical} where we introduce and discuss the predicted Baldwin–Phillips–Terlevich (BPT) diagrams \citep{baldwin81}.

Unlike to \ha\ and \hb, the metastable and electronic lines of [S\,{\sc ii}], [N\,{\sc ii}], and [O\,{\sc iii}] are primarily excited by collisions of S$^+$, N$^+$, and O$^{++}$ with thermalized electrons. Because the excited levels lie at $\sim 2.5 \times 10^4$ K above the ground state, the corresponding emissivity not only depend on the ionization profiles of the cloud but also on its temperature, and, in particular, on the amount of gas above $\sim 5000$ K. The integrated intensities of all these lines are therefore built up at the border of the cloud where the gas is both ionized and warm. Since the ionization potential of O$^+$ is considerably larger than that of O, N, or S, [O\,{\sc iii}] lines necessarily trace the outskirt of the 
[O\,{\sc ii}], [N\,{\sc ii}], and [S\,{\sc ii}] emitting regions. In this framework, the dependencies on the input parameters are straightforward and simply follow the results presented in the previous section. Low density or strong X-ray radiation field naturally favor large abundances of ionized species at the border of the cloud and the depth over which the gas is at high temperature. The integrated intensities of [O\,{\sc ii}], [N\,{\sc ii}], [S\,{\sc ii}], and [O\,{\sc iii}] lines therefore increase with $G_{\rm X}$ but also with \vtur\, if \vtur $\geqslant$ \vtur$^{\rm lim}$ (Figs. \ref{fig:exproplx2} \& \ref{fig:exproplx2tur}).

All the lines described above are tracers that carry only a small fraction $\sim 10^{-2}$ of the input X-ray energy flux. Indeed, because of the assumed initial SED, the impinging X-ray radiation field is preferentially reprocessed into the fine structure lines of [O\,{\sc i}] and [C\,{\sc ii}], providing that $A_{\rm V}$ is sufficiently large. As a result, for most of the models considered here, these lines not only carry a substantial amount of the input radiative flux, but also strongly depend on the size of the cloud. Evidently, the integrated intensities of both [O\,{\sc i}] and [C\,{\sc ii}] increase with $G_{\rm X}$. The increase of \vtur, instead, has a different effect on the two cooling lines. When \vtur\, increases, the density decreases while the temperature increases. For large densities, C+ is converted into C very early in the cloud (see Fig. D1, left), while the transition happens later in a low density medium (see Fig. D1, right). Thus, the integrated intensities of [C\,{\sc ii}] increase with \vtur. The integrated intensities of [O\,{\sc i}], instead, slightly decrease due to the increasing temperature.



\section{Comparison with the observables}\label{sec:results}
In this section the predicted line emission from our grid of models are compared with the observations, from optical to submillimeter, of several BCGs.

\subsection{Optical tracers}\label{sec:optical}
The combination of the line ratios \oiiiop/\hb, \niiop/\ha\, (\siiaop+\siibop)/\ha\, and \oiop/\ha\, are commonly used in the so-called BPT diagrams to identify the excitation mechanism of the optical line-emitting gas. \cite{mcdonald12} used this technique to investigate the contributors to the ionization of the filaments in nine cool-core galaxy clusters. Overlapping on the BPT diagrams grid of models from previous studies to the observed values, the authors eliminated most of the heating mechanisms suggested. 
We compare our grid of models with the observed values of their BCG sample\footnote{We reproduce the BPT diagrams of the nine cool core galaxy clusters reporting all of the individual data for each object.} in Fig.~\ref{fig:bpt}. To make easier the reading of these plots, we show in the Appendix~\ref{sec:retiovsav}, as examples, the BPT diagrams varying Z and \gx\, for a fixed turbulence of 10 km s$^{-1}$ (Figure~\ref{fig:bpt_1}), and those varying only the turbulence and the metallicity for single values of \gx = 100 (Figure~\ref{fig:bpt_2}). These figures show the line ratios predicted at different \av: 0.001 mag (top row), 0.1 mag (second row), 1 mag (third row) and 5 mag (bottom row). 

In Fig.~\ref{fig:bpt} we can immediately see that, in general, our grid of models can reproduce the observables except the \oiop/\ha\ ratio. The observed values in the [N\,{\sc ii}]-diagram are well reproduced at all of the explored \av. Note that varying \av\, the best model that reproduces the specific observed ratio changes. Also the predicted (\siiaop+\siibop)/\ha\, are in agreement with the data, even if increasing \av\, increases the predicted (\siiaop+\siibop)/\ha\, ratio. The predicted \oiop/\ha, instead, are in agreement with the observables only for very low \av\, (\av\,=\,10$^{-3}$ mag). 
Table~\ref{tab:BPTsummary} summarizes which is, among all of the models, the restricted parameter space that can reproduce the optical observations. The range of \gx\, values that can reproduce most of the observables, for any metallicity, moves to lower values with the increasing of the turbulence and higher values with the increasing of \av. The different position of the models on the BPT-diagrams highlights the impact of \gx, turbulence and \av\, on the physical properties (i.e., temperature, electron fraction, density) of the cloud and thus, affect the ratio of the optical line ratios. 

The optical line ratios of the BPT-diagrams trace different sections of the cloud. As described in the previous section, \oiiiop\, arises from the outskirt of the environment compared to \niiop, and [S\,{\sc ii}] lines, while \oiop, which traces warm dense gas, comes from the ionized front. Thus, varying \av\, affects the line ratios for a given combination of \gx\, and turbulence. For those models with low excitation sources (low \gx\, and turbulence) the cloud does not have an ionized phase, thus there is no impact of the variation of \av\, on the line ratios. The increase of \gx\, and/or the turbulence moves the ionized front deeper into the cloud. Thus, for those models with excitation sources strong enough to ionized the gas, varying \av\, changes the relative phase distribution of the clouds, and consequently the predicted optical line ratios vary. In particular the increasing of the over-estimation of \oiop/\ha\ with \av, suggests that the observed optical emission arise from a cloud with low \av, in other words, a large fraction of the gas in the filaments can be reproduce by a matter bounded cloud.

We note that the observed \oiiiop/\hb\, ratio given by \cite{mcdonald12} can be very small (0.1-0.03). This was already discussed in \cite{hatch06} who noticed the unexpected complete lack of \oiiiop\, in most of the regions inside the filaments around NGC\,1275. The observed optical line ratios may have several excitation mechanisms, such as star-formation and particle excitation, which may be varying the \oiiiop/\hb\, line ratio. The small values in \cite{mcdonald12} correspond to regions where there is not evidence of star-formation. Powering the optical nebula without strong \oiiiop\, line can be explained if cosmic ray heating/ionization is invoked as described by \cite{ferland09}. Another solution is the strategy adopted in this work, i.e. use an EUV continuum that can reproduce the observed optical line ratios (see Sec.~\ref{sec:SEDshape}). We leave for a future work the detailed comparison of the current model predictions with the observed emission lines in the filaments of NGC\,1275.

\begin{figure*}[htbp!]
\centering
        \includegraphics[trim={0cm 1.35cm 0cm 0.7cm}, clip, width=.95\hsize]{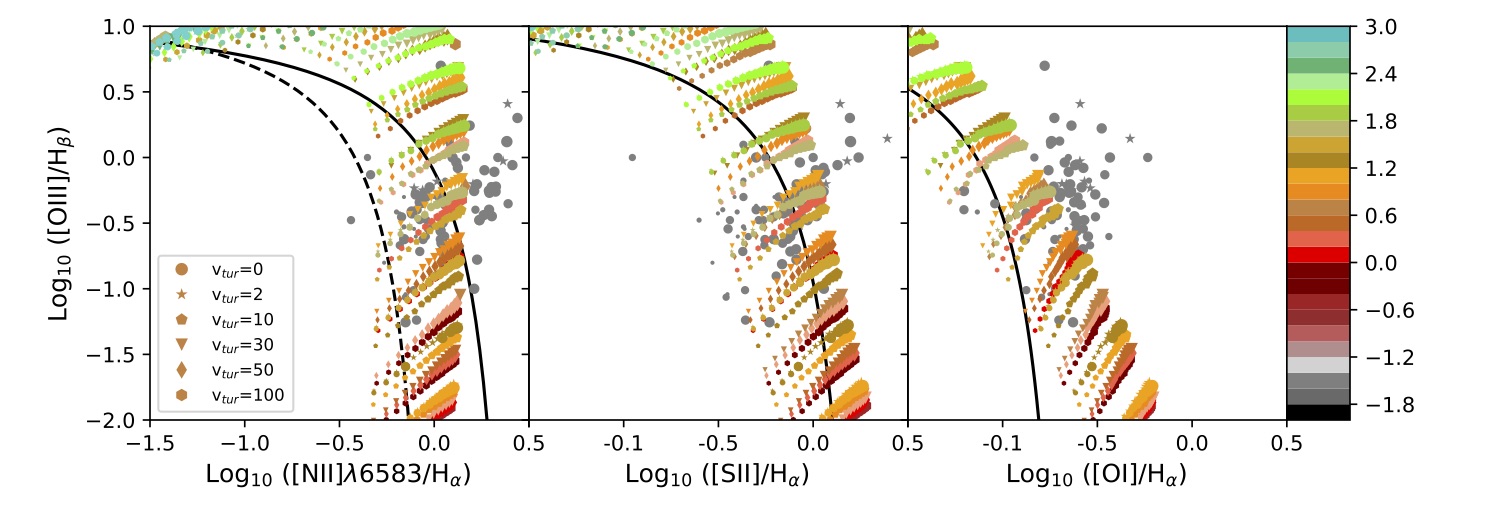}
        \includegraphics[trim={0cm 1.35cm 0cm 0.25cm}, clip, width=.95\hsize]{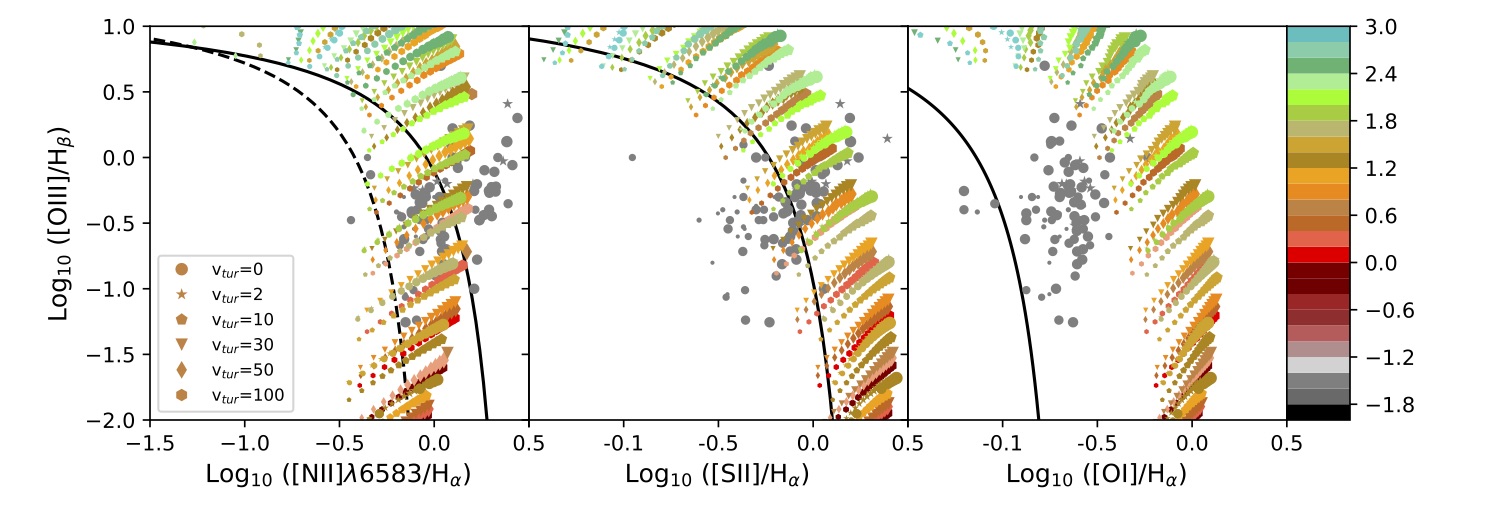}
        \includegraphics[trim={0cm 1.35cm 0cm 0.25cm}, clip, width=.95\hsize]{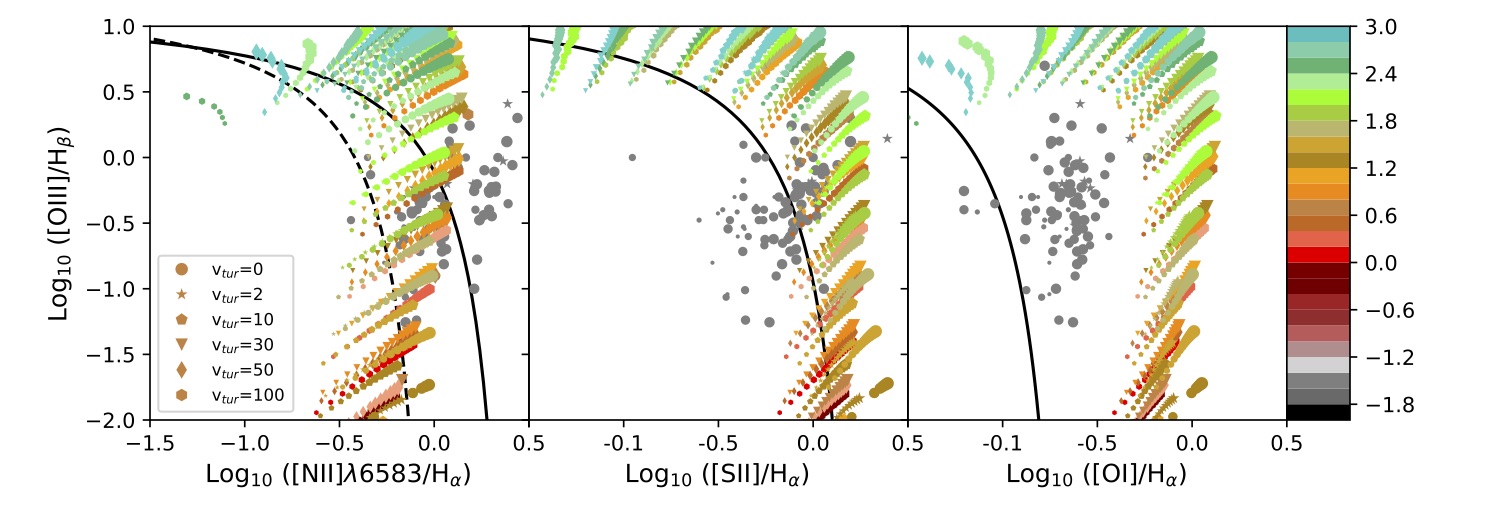}
        \includegraphics[trim={0cm 1.35cm 0cm 0.25cm}, clip, width=.95\hsize]{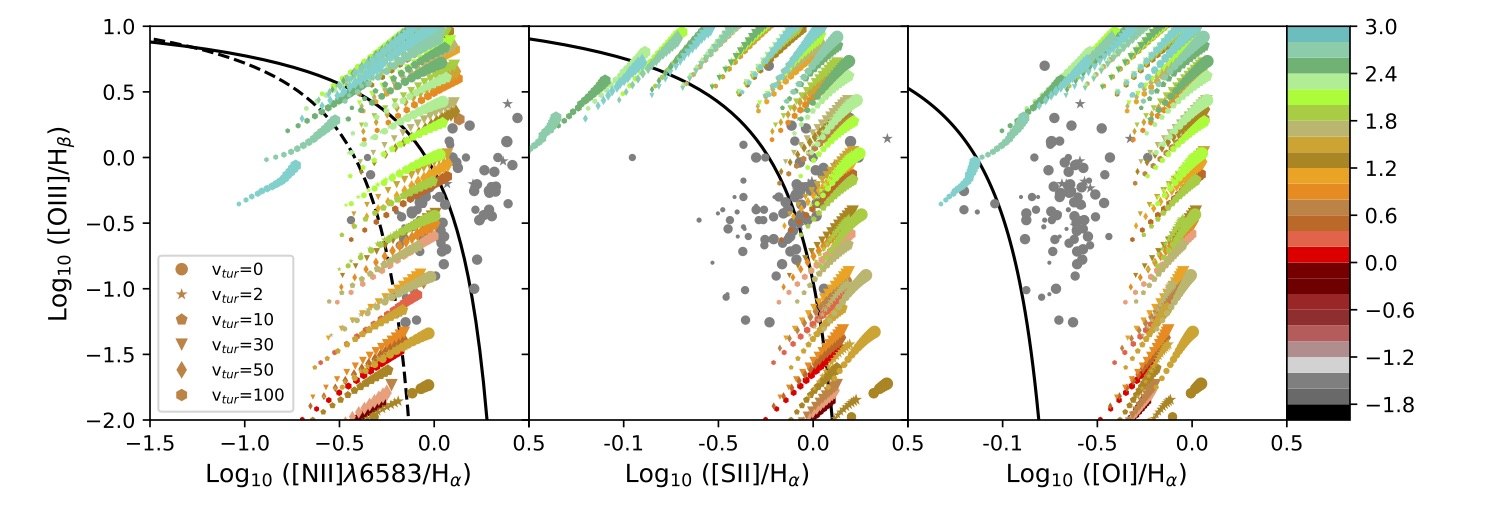}
        \caption{BPT diagrams. The gray dots (filament values) and the stars (nucleus values) represent the data from \cite{mcdonald12}. The size of the dots decreases with the increasing of the distance from the nuclei of the corresponding object. The solid black line is the upper limit for \hii\, regions by \cite{kewley01} and the dashed black line is the lower limit for AGN by \cite{kauffmann03}. The predicted cumulative line ratios from the models are overlapped. The color bar correspond to the different values of \gx\, (the logarithm of the value is written on the bar); each symbol correspond to one turbulence value and the size of the symbols increase with the metallicity (from 0.3 to 1 Z$_{\odot}$). {\it Top row}: cumulative emission at \av\ = 0.001 mag; {\it second row}: cumulative emission at \av\ = 0.1 mag; {\it third row}: cumulative emission at \av\ = 1 mag; {\it Bottom row}: cumulative emission at \av\ = 5 mag. {\it Left column}: \oiiiop/\hb\ vs. \niiop/\ha; {\it central column}: \oiiiop/\hb\ vs. [S\,{\sc ii}]/\ha; {\it right column}: \oiiiop/\hb\ vs. [O\,{\sc i}]/\ha.}
    \label{fig:bpt}
\end{figure*}

\begin{table*}[htbp!]
\centering
     \caption{Restricted ranges of free parameters are given for 4 values of \av\,= 0.001, 0.1, 1 and 5 mag.} 
        {\small \begin{tabular}{r r c c r c c r c c r c c}
       		\hline
       		\hline
        		 \noalign{\smallskip}
       			Turbulence & \av\ & \gx\,& Z & \av\ & \gx\,& Z & \av\ & \gx\,& Z & \av\ & log$_{10}$\gx\,& Z\\
			$[$km s$^{-1}$] & [mag] &  & Z$_{\odot}$ & [mag] &  & Z$_{\odot}$ & [mag] &  & Z$_{\odot}$ & [mag] &  & Z$_{\odot}$\\
		 \noalign{\smallskip}
		 \hline
		\noalign{\smallskip}
		0 & 0.001 & 10$^{1.2}$ - 10$^{2}$ & all & 0.1 & 10$^{1.6}$ - 10$^{2.6}$ & all & 1 & 10$^{1.6}$ - 10$^{3}$ & all & 5 & 10$^{1.6}$ - 10$^{3}$ & all \\
		2 &  & 10$^{1.2}$ - 10$^{2}$ & all & & 10$^{1.6}$ - 10$^{2.6}$ & all &  & 10$^{1.6}$ - 10$^{3}$ & all &  & 10$^{1.6}$ - 10$^{3}$ & all\\
		10 &  & 10$^{1}$ - 10$^{2}$ & all & & 10$^{1.4}$ - 10$^{2.4}$ & all & & 10$^{1.6}$ - 10$^{2.8}$ & all &  & 10$^{1.6}$ - 10$^{3}$ & all \\
		30 &  & 10$^{0.6}$ - 10$^{1.4}$ & all & & 10 - 100 & all & & 10 - 10$^{2.2}$ & all &  & 10 - 10$^{2.4}$ & all \\
		50 &  & 10$^{0.2}$ - 10$^{1}$ & all & & 10$^{0.6}$ - 10$^{1.6}$ & all & & 10$^{0.6}$ - 10$^{1.8}$ & all &  & 10$^{0.6}$ - 10$^{2.2}$ & all \\
		100 &  & 10$^{-0.2}$ - 10$^{0.6}$ & all & & 1 - 10$^{1.2}$ & all & & 10$^{0.2}$ - 10$^{1.4}$ & all &  & 1 - 10$^{1.6}$ & all \\	
	 \noalign{\smallskip} 
	 \hline
	 \hline
\end{tabular}}
\\
 \label{tab:BPTsummary}
\end{table*}

\subsection{Infrared tracers}
Unlike the optical lines, infrared observations are not affected significantly by extinction and they cover a wide range of ionization potentials and critical densities. Moreover, line ratios of infrared lines are almost independent from the temperature. 

\subsubsection{\neii/\neiii}
The line ratio \neii/\neiii, is a perfect tracer of the intensity of the radiation source, since the two lines are emitted by the same element but with different ionization stages. The \neiii\ has higher ionization potential than \neii, 41 eV and 21.6 eV, respectively, and they have similar critical density, 3$\times$10$^{5}$ cm$^{-3}$ for \neiii\ and 7$\times$10$^{5}$ cm$^{-3}$ for \neii. Both lines are found exclusively in \hii\, regions, with \neiii\, arising from the layer closer to the radiative source compared to \neii. These two lines have been detected for several BCGs with the IRS instrument onboard of {\it Spitzer}. We collected the observed ratio [Ne\,{\sc ii}]/[Ne\,{\sc iii}] for a total of 10 BCGs (values from \citealt{donahue11}, \citealt{egami06}, and \citealt{johnstone07}) and compare them with the predicted line ratio from our grid of models. The comparison is shown in Figure~\ref{fig:IRratios}, top row. The observed ratio of the full sample covers the range between 1.10 and 5.25, and it is represented by the blue background. We do not present the theoretical line ratio for all of the models, but only for few representative cases, with the aim of understanding how the ratio changes as a function of the free parameters: \gx, turbulence, metallicity and \av. On the left panel we fix v$_{\rm tur}$ and Z, and we vary only \gx. For low \gx, the model can not reproduce the observables.  Increasing the intensity of the radiation field the model produces more energetic photons, boosts \neiii\, and the ratio decreases, with lower values at low \av. After the ionization front, the ratio becomes constant. Similar behavior is shown varying only the turbulence (central panel). We see that the combination of \gx=\,10 and v$_{\rm tur}$=\,0-2 km s$^{-1}$ can reproduce the observed range at every \av. The increasing of v$_{\rm tur}$ boosts \neiii\ and moves the predicted values that can reproduce the observations inside the cloud. The consequence of the variation of metallicity on the [Ne\,{\sc ii}]/[Ne\,{\sc iii}] ratio is almost negligible and all of the model with v$_{\rm tur}$\,=\,10 km s$^{-1}$ and \gx=10 can reproduce the observed ratio (right panel). The restricted range of the parameter space that can reproduce the observations taking into account all of the models is presented in Table~\ref{tab:IRsummary}. The models with \gx\ between 1 to 10$^3$ can reproduce the line ratio for an extended range of \av. The smallest value of \av\, which match observed [Ne\,{\sc ii}]/[Ne\,{\sc iii}] ratio varies with the combination of \gx\, and v$_{\rm tur}$. The variation of the metallicity affects slightly the behavior of the predicted line ratio. The impact of the metallicity on the range of \av\, is indicated in Table~\ref{tab:IRsummary} with different letters. 

\subsubsection{\oia/\cii}
\begin{figure*}[htbp!]
        \includegraphics[trim={0cm 1cm 0cm 0cm}, clip, width=\hsize]{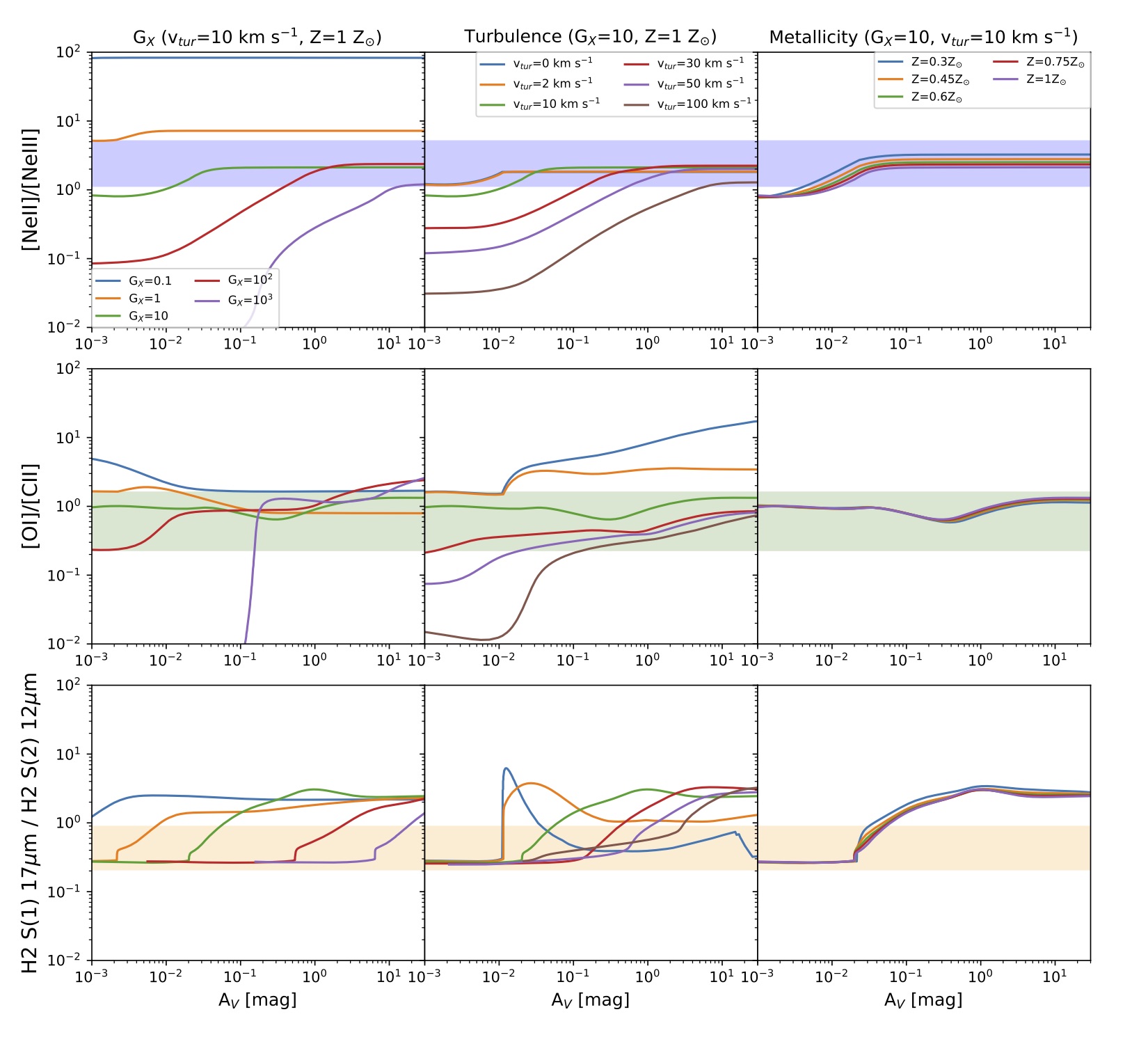}
        \caption{Comparison of observed and predicted cumulative line ratios: [NeII]/[NeIII] (top), \oia/\cii\,(middle) and H$_{2}$ S(1) 17.0 $\mu$m/H$_{2}$ S(2) 12.3 $\mu$m (bottom), as a function of \av. The blue background highlights the observed line ratio of [NeII]/[NeIII], range of values between 1.10 and 5.25 (data from \citealt{donahue11}, \citealt{egami06}, and \citealt{johnstone07}). The observed line ratio \oia/\cii\, range, between 0.22 to 1.64 (values from \citealt{edge10}, \citealt{mittal11}, \citealt{mittal12}, and \citealt{werner14}), is shown in green, and the yellow background highlights the observed ratio of the two pure rotational H$_{2}$ lines, between 0.2 and 0.9 (values from \citealt{donahue11}, \citealt{egami06}, and \citealt{johnstone07}). The column on the left shows the predicted ratios for models with v$_{\rm tur}$\,=\,10 km s$^{-1}$ and solar metallicity, the central column the ratios for models with \gx=10 and solar metallicity and the column on the right the ratios for models with v$_{\rm tur}$\,=\,10 km s$^{-1}$ and \gx=10.}
    \label{fig:IRratios}
\end{figure*}

The far-infrared line ratio \oia/\cii\, observed by {\it Herschel} is a good tracer of the electron density if the lines are optically thin, since the critical densities of these two lines are different. The critical density for \oia\ is $\sim$5$\times$10$^5$ cm$^{-3}$, and for \cii\ is $\sim$3$\times$10$^3$ cm$^{-3}$, much higher. The range of observed \oia/\cii\ line ratios (0.22 - 1.64) is shown, in the central row of Figure~\ref{fig:IRratios}, as the green background (values from \citealt{edge10}, \citealt{mittal11}, \citealt{mittal12}, and \citealt{werner14}). Our models match those observations for a large range of free parameters, as summarized in Table~\ref{tab:IRsummary}. Since Carbon and Oxygen abundances are scaled by the same factor, the metallicity has very little impact on this line ratio, while the variation of \gx\, and turbulence has a clear consequences on the profile of the \oia/\cii\ line ratios. 
For a fixed value of \gx, the range of \av\ that fit the data increases with the increasing of the turbulence. The same behaviour can be observed for models with a fixed value of v$_{\rm tur}$ and different \gx, i.e. increasing \gx\, moves the minimum value of the range of \av\, that fit the data to higher \av. 

\begin{table*}[htbp!]
\centering
     \caption{Restricted parameter space of the free parameters that reproduce the infrared observations.} 
        {\small \begin{tabular}{l | r r c c |  r r c c}
       		\hline
       		\hline
        		 \noalign{\smallskip}
       			Line or ratio & Turbulence & \gx & \av\ & Z & Turbulence & \gx & \av\ & Z\\
			& $[$km s$^{-1}$] & & [mag] & & $[$km s$^{-1}$] & & [mag] & \\
		 \noalign{\smallskip}
		 \hline
		\noalign{\smallskip}
		$[$NeII]/[NeIII] & 0 & 10 & all & A & 30 & 1 & all & A\\
		&  & 100 & $\geq$0.1 & B &  & 10 & $\geq$0.1 & B\\
		&  & 10$^3$ & $\geq$7 & B &  & 100 & $\geq$0.3 & B \\
		 \hline
		$[$OI]/[CII] &  & 10 & $\leq$0.01 & A &  & 0.1, 1, 10 & all & A\\
		 &  &  &  & &  & 100 & $\geq$0.03 & B\\
		  &  &  & & &  & 10$^3$ & $\geq$0.4 & B\\
		  \hline
		H$_{2}$ 0-0 S(1)/ & & 0.1  & $\leq$0.02 & C &  & 0.1 & $\leq$0.003 & F\\
		H$_{2}$ 0-0 S(2) & & 1 & $\geq$0.08 & A & & 1 & $\leq$0.03 & F\\
		& & 10 & $\leq$0.01 and $\geq$0.03 & D & & 10 & $\leq$0.3 & F \\
		& & 100 & $\leq$0.2 and $\geq$1 & C & & 100 & $\leq$7 & F\\
		& & 10$^3$ & $\leq$3 and $\geq$20 & A & & 10$^3$ & $\geq$0.3 & F \\
		 \noalign{\smallskip}
		 \hline
		  \hline
		\noalign{\smallskip}
		$[$NeII]/[NeIII] & 2 & 10 & all & A & 50 & 0.1 & all & A\\
		&  & 100 & $\geq$0.2 & B &  & 1 & $\geq$0.02 & B\\
		&  & 10$^3$ & $\geq$10 & B &  & 10 & $\geq$0.5 & A\\
		 \hline
		$[$OI]/[CII] &  & 10 & $\leq$0.01 & A &  & 0.1, 1 & all & B\\
		&  & 100 & $\leq$ 0.2 & E & & 10 & $\geq$0.02 & B \\
		&  & 10$^{3}$ & between 0.07 and 2 & B &  & 100 & $\geq$0.01 & B\\
		&  & & & &  & 10$^3$ & $\geq$1 & B\\
		 \hline
		H$_{2}$ 0-0 S(1)/ & & 10 & $\leq$0.01 & A & & 0.1 & $\leq$0.003 & F\\
		H$_{2}$ 0-0 S(2) & & 100 & $\leq$0.2 & E & & 1 & $\leq$0.04 & F\\
		& & 10$^3$ & $\leq$4 & A & & 10 & $\leq$1 & F\\
		& & & & & & 100 & between 0.07 and 10 & F\\
		& & &  & & & 10$^3$ & $\geq$1 & E\\
		 \noalign{\smallskip}
		 \hline
		  \hline
		\noalign{\smallskip}
		$[$NeII]/[NeIII] & 10 & 1 & $\leq$0.003 & F & 100 & 0.1 & all & A\\
		&  & 10 & $\geq$0.01 & B &  & 1 & $\geq$0.1 & B\\
		&  & 100 & $\geq$0.3 & B &  & 10 & $\geq$3 & B \\
		&  & 10$^3$ & $\geq$10 & B &  &  &  & \\
		 \hline
		$[$OI]/[CII] &  & 0.1 & $\geq$0.02 & B &  & 0.1 & $\leq$0.01 and $\geq$0.7 & A\\
		 &  & 1 & all & A &  & 1 & $\geq$0.02 & B\\
		 &  & 10 & all & A &  & 10 & $\geq$0.1 & B\\
		 &  & 100 & $\leq$3 & A &  & 100  & $\geq$0.3 & B\\
		 &  & 10$^{3}$ & between 0.01 and 10 & B &  & 100 & $\geq$3 & B\\
		 \hline
		H$_{2}$ 0-0 S(1)/ & & 1 & $\leq$0.008 & F & & 0.1 & $\leq$0.01 & F\\
		H$_{2}$ 0-0 S(2) & & 10 & $\leq$0.05 & F & & 1 & $\leq$0.2 & F\\
		& & 100  & $\leq$3 & F & & 10 & between 0.02 and 3 & F\\
		& & 10$^3$ & $\leq$20 & F & & 100 & $\geq$0.3 & F\\
		& &  & & & & 10$^3$ & $\geq$3 & B\\
	 \noalign{\smallskip} 
	 \hline
	 \hline
\end{tabular}}\\
{\small{\bf Notes.} A = There is not difference in the range due to the variation of Z.\\ 
B = The minimum value of \av\, decreases with the decreasing of Z.\\
C = The minimum value of \av\, increases with the decreasing of Z. \\
D = The range of \av\, that can reproduce the observed values moves to lower \av\, for lower Z.\\
E = The maximum value of \av\, increases with the decreasing of Z.\\
F = The maximum value of \av\, decreases with the decreasing of Z.}
\\
 \label{tab:IRsummary}
\end{table*}

\subsubsection{Pure rotational H$_{2}$ lines}
In the infrared we can also detect the line emission of the warm molecular gas ($\geq$100 K) emitted by the rotational and vibrational H$_{2}$ transitions. 
In the bottom row of Figure~\ref{fig:IRratios}, we compare the observed H$_{2}$ pure rotational line ratio S(1)17$\mu$m/S(2)12$\mu$m (values from \citealt{donahue11}, \citealt{egami06}, and \citealt{johnstone07}) to our model predictions. The observed line ratio is between 0.2 and 0.9, represented by the orange background, and the predicted line ratios from representative models are shown as a function of \av. At a moderate constant turbulent heating rate (\vtur\,=\,10 \km, left panel), because the S(1) line has a lower excitation temperature than the S(2) line, all models show an increase of the H$_{2}$ S(1)/S(2) line ratio with \av, except for very low \gx\, where it is roughly constant (in that case the cloud temperature is almost constant at \av > 0.01). We note that the models match the observed range of ratios where the medium is mostly atomic. For a fixed \gx=10 and solar  metallicity (middle panel), the observed ratio can be reproduced by all of the turbulence values, but for a different range of \av. Finally, the metallicity affects only slightly the behavior of the predicted line ratio (right panel). The summary of the combination of free parameter values that can reproduce the observed ratio H$_{2}$ S(1)~17.0~$\mu$m/S(2)~12.3~$\mu$m is provided in Table~\ref{tab:IRsummary}. 

\subsubsection{Ro-vibrational H$_{2}$ lines}
\begin{figure*}[htbp!]
        \includegraphics[width=\hsize]{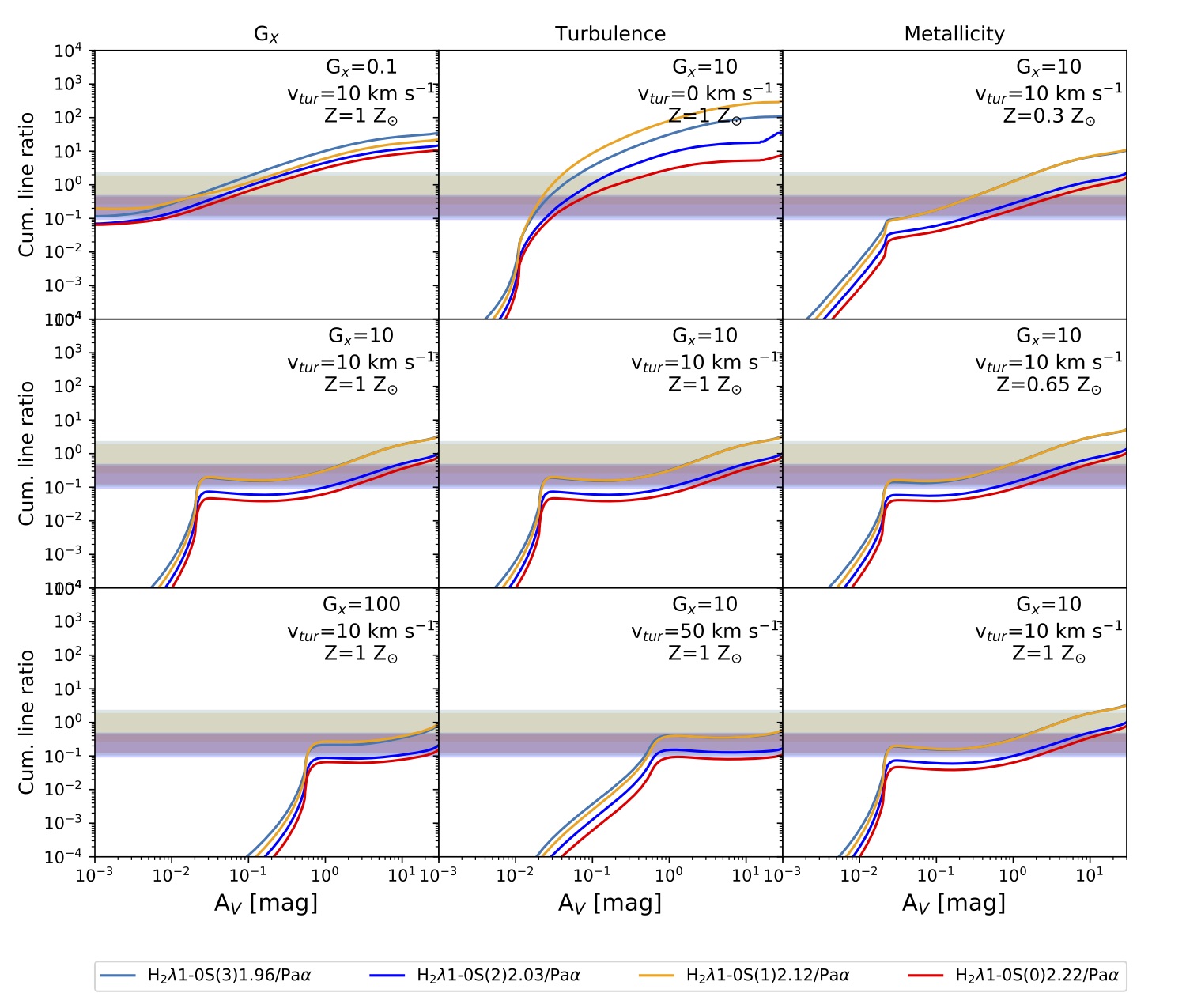}
        \caption{Comparison of observed and predicted ro-vibrational H$_{2}$ lines, normalized by Pa$\alpha$, as a function of \av\, for the models described in Sect.~\ref{sec:physicsmodels}. The different background highlights the observed ratios, between 0.1 and 2.3 (values from \citealt{edge02}). For each ratio, the color of the background is the same of that of the theoretical cumulative ratio. The column on the left shows the predicted cumulative emission for models with turbulence of 10 km s$^{-1}$, solar metallicity and various values of \gx: 0.1 (top), 10 (middle) and 100 (bottom); the central column shows the predicted cumulative emission for models with \gx=10, solar metallicity and various values of turbulence: 0 (top), 10 (medium) and 50 (bottom) km s$^{-1}$; the column on the right shows the predicted cumulative emission for models with turbulence 10 km s$^{-1}$, \gx=10 and various values of metallicity: 0.3 (top), 0.65 (middle) and 1 (bottom) Z$_{\odot}$.}
    \label{fig:rovib}
\end{figure*}

Also the ro-vibrational H$_{2}$ lines that emits $\sim$2 $\mu$m have been observed in several BCGs, and presented by \cite{edge02}. In Figure~\ref{fig:rovib} we compere the observed values of H$_{2}$ 1-0 S(3), H$_{2}$ 1-0 S(2), H$_{2}$ 1-0 S(1) and H$_{2}$ 1-0 S(0), normalized by Pa$_{\alpha}$, to our predicted H$_{2}$/Pa$_{\alpha}$, ratios. In our models, these line emission are excited by the collision with the secondary electrons produced by the interaction of the X-ray radiation field with the medium. For most of our models, this excitation mechanism is enough to produce the H$_{2}$ emission inside the cloud and we do not need to add cosmic rays. Increasing the X-ray radiation field intensity (left column) the peak of the emission moves to higher \av. Consequently the cumulative intensity of the H$_{2}$ lines reaches higher values deeper into the cloud. However, the increasing of \gx\, decreases the density of the cloud at low \av, thus the cumulative emission profiles of these lines become steeper. The variation of the turbulence, instead, do not changes the values of the secondary electron rate (central column), but it affects the structure of the cloud (see sec.~\ref{sec:physicsmodels}). The combination of these two factors pushes the intensity of the ro-vibrational lines down for higher turbulence. Finally, with the increasing of the metallicity (right column) the line emission decrease. All of the models that we show in Fig.~\ref{fig:rovib} can reproduce the normalized ro-vibrational line emission, even if at different \av.

\subsection{Sub-millimeter lines}\label{sec:co}
\begin{figure}[htbp!]
        \includegraphics[trim={0.5cm 0.5cm 0.5cm 0cm}, clip,width=\hsize]{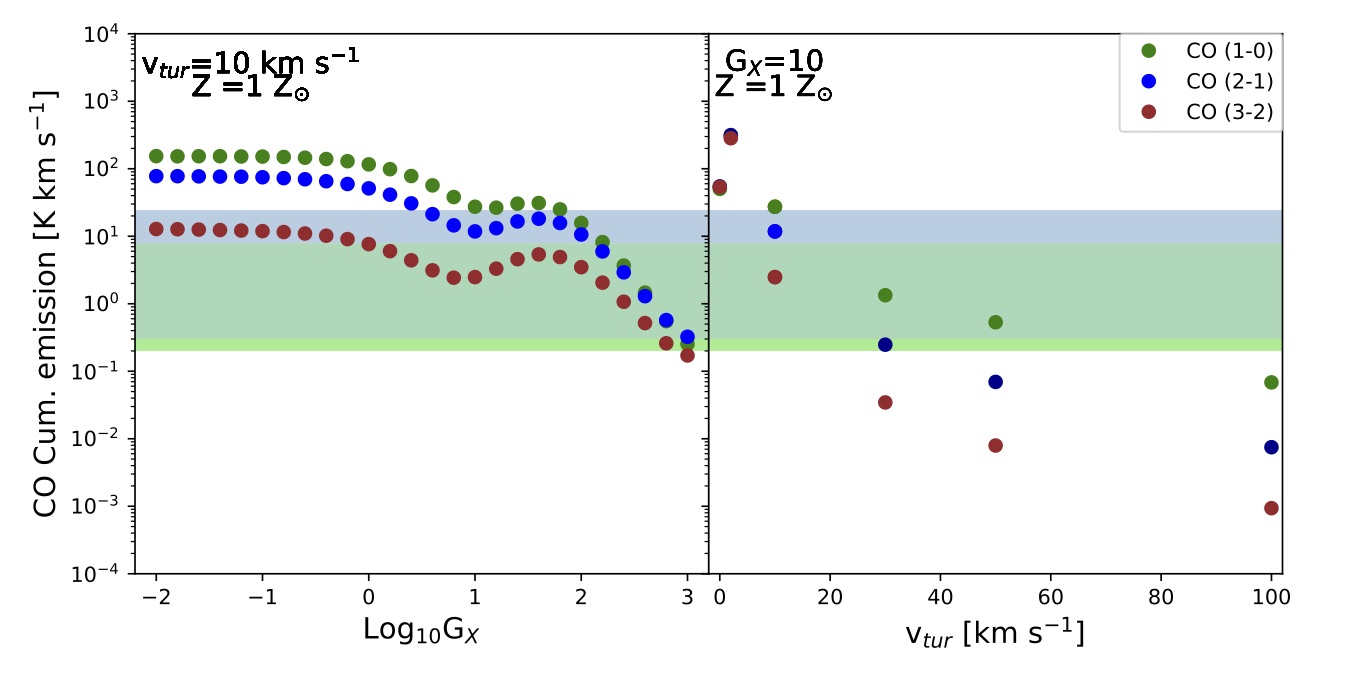}
        \caption{Comparison of observed (green and blue background) and predicted cumulative CO emission: 1-0, 2-1 and 3-2. The predicted values are for \av=30 mag (end of the cloud). The panel on the left shows the predicted values as a function of \gx, for models with turbulence of 10 \kms\, and solar metallicity. The panel on the right shows the predicted values as a function of turbulence, for models with \gx=10 of and solar metallicity.}
    \label{fig:mol}
\end{figure}

Finally, we compare the predicted CO(1-0), (2-1), (3-2) transitions, which trace the cold ($\sim$10K) molecular gas, to the observations presented in \cite{salome08}. Figure~\ref{fig:mol} shows the observed range of CO (1-0) and CO (2-1) with a green and a blue background, respectively, and the predicted values of the three transitions for models with fixed turbulence and metallicity but different values of \gx (left panel), and for models with fixed \gx\, and metallicity but varying the turbulence (right panel). Since the cold molecular gas traced by the CO transitions is expected to arise from the deepest region of the cloud, we compared the observations only with the predicted cumulative emission at \av=30 mag. For a fixed v$_{\rm tur}$=10 \kms\, and Z=Z$_{\odot}$, all the models with \gx\,$\geq$10 can reproduce the observed CO(1-0) and all the models with \gx\,$\geq$1.6 can reproduce the observed emission CO(2-1). While, fixing Z=Z$_{\odot}$ and \gx=10, only the model with v$_{\rm tur}$ = 10 and 30 \kms\, can reproduce both transitions. 
These comparisons restrict drastically the `good' parameter space, in particular the turbulence. We would like to remind that these results are based on a single model component and a more complex multi-component model could bring different results. An additional limitation in these models is that the turbulence is treated as a constant heating rate everywhere in the cloud, which is unrealistic.

\section{Origin of the soft X-ray/EUV emission}\label{sec:discussion}
The actual model assumes that the soft X-ray and the EUV photons that power the nebula are produced by the cooling of the hot ICM in the filaments. In this section, we investigate whether the input X-ray intensity of our models can power the total line intensities, and we compare with observational estimates from the Perseus cluster.

To give orders of magnitude, a \gx=10 corresponds to an input X-ray surface brightness in the range 0.6-3.4 keV of $\sim$4 $\times$ 10$^{-14}$ erg cm$^{-2}$ s$^{-1}$ arcsec$^{-2}$. For a typical filament in Perseus of 27''$\times$1'' at the distance of 80 Mpc, this means a luminosity of $\sim$1$\times$10$^{42}$ erg s$^{-1}$.  
\citealt{sanders07} analyzed the X-ray emission in a region of 27.3 $\times$ 1 arcsec$^{2}$. They computed that the difference in temperature of the cooling X-ray emitting gas from the surrounding gas was $\sim$2.7 keV. This imply a released energy of $\sim$2.5$\times$10$^{54}$ erg if the X-ray emitting filament was to cool out of the surrounding ICM.  Assuming a timescale for this process of $\sim$10$^{5}$ yr, they conclude that the gas cooling provides enough energy to power about 10$^{42}$ erg/s.

\begin{figure}[htbp!]
        \includegraphics[trim={0cm 0cm 0.5cm 0cm}, clip,width=\hsize]{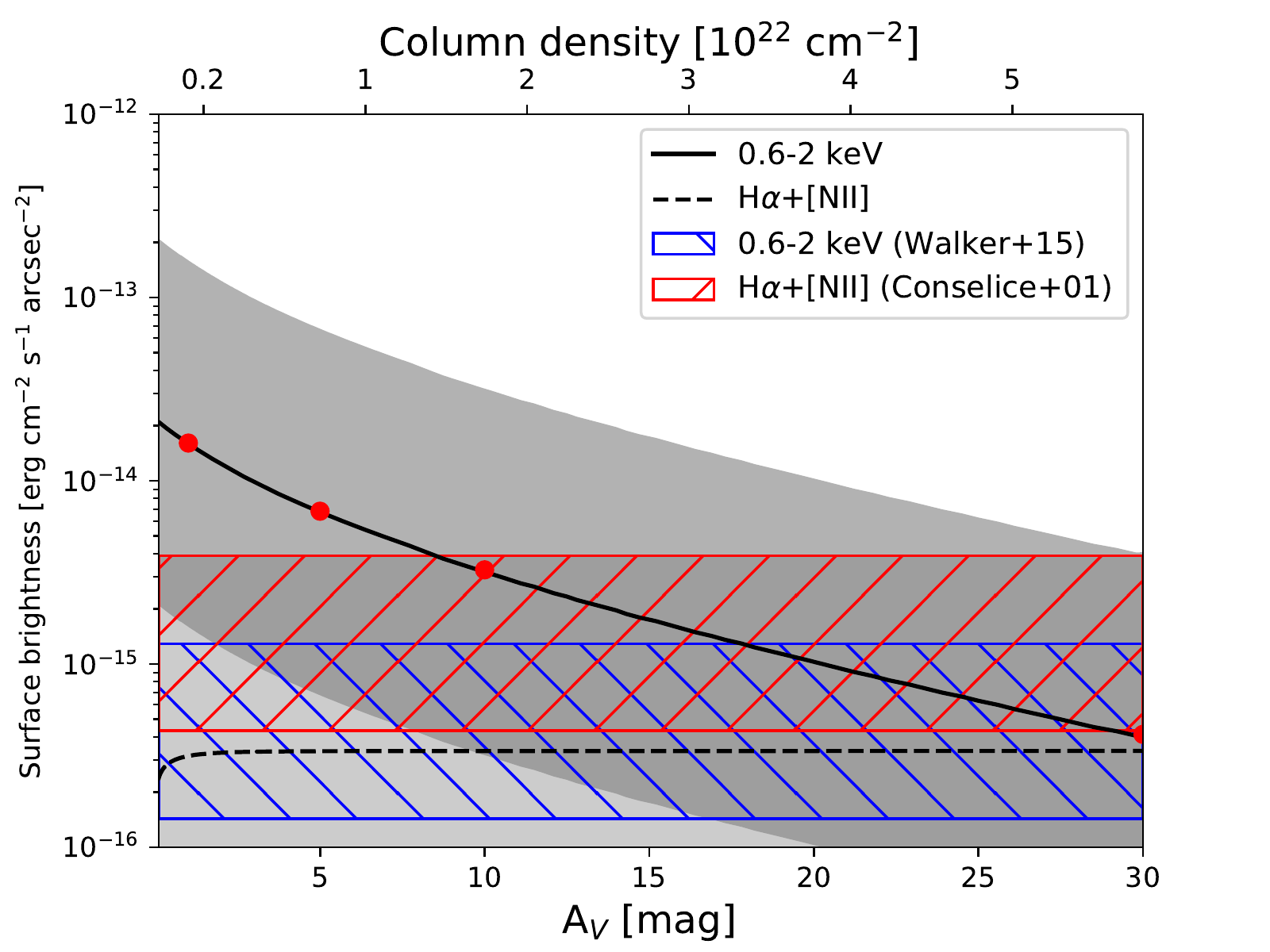}
        \caption{Predicted cumulative surface brightness of \ha+\niiopab\, (dashed line) and predicted ionizing continuum 0.6-2 keV X-ray band (solid line) as a function of \av\, for a model with \gx=10, v$_{tur}$=10 \kms\, and Z=Z$_{\odot}$. The dark grey and soft grey background show the range of predicted ionizing continuum 0.6-2 keV X-ray and cumulative surface brightness of \ha+\niiop,  respectively, for models with \gx\, in the range 1 - 100. The red dots show the predicted transmitted continuum 0.6-2 keV X-ray band estimated at \av\,= 1, 5, 10, 30 mag. The area covered by red (blue) diagonal lines represent a factor of three of the observed surface brightness of \ha+\niiopab\, from \cite{conselice01} for the region 15 of the paper, 1.3$\times$10$^{-15}$ \ergcms arcsec$^{-2}$ (of 0.6-2 keV X-ray band from \cite{walker15}, 4.3$\times$10$^{-16}$ \ergcms arcsec$^{-2}$).}
    \label{fig:haXray}
\end{figure}

The authors also discuss the fact that the X-ray surface brightness in 0.5-2 keV is almost similar to the \ha+\niiop\ surface brightness given by \cite{conselice01} (Fig.18 of \citealt{sanders07}). In order to account for the total reprocessed emission in all the different lines, they estimate that the input power must be a factor of 20 larger than the one required for the  \ha\ only. This leads to a required input power of $\sim$10$^{42}$ erg/s for a typical filament in the Perseus cluster. The present models show that the predicted surface brightness in the range 0.6-2 keV is similar to the predicted \ha+\niiopab\, surface brightness for \av\,$>$20, which corresponds to hydrogen column density $\sim$10$^{22}$ cm$^{-2}$. A cloud with large optical depths is needed to reprocess the X-rays and excite the infrared H$_2$ lines (Figure~\ref{fig:rovib}). As an example, Figure~\ref{fig:haXray} shows the behavior of the two emissions for the model with \gx=10, v$_{\rm tur}$=10 \kms\,and Z=Z$_{\odot}$. The predicted surface brightness of the reprocessed soft X-ray from the model is comparable with the observed value, 4.3$\times$10$^{-16}$ \ergcms arcsec$^{-2}$, within a factor of three for \av$\geq$18 mag. The \ha+\niiopab\, surface brightness, instead, is under-predicted by a factor of $\sim$4. It is important to remind that large uncertainties are related to the observations, such as the extinction correction for both Galactic and internal reddening (e.g. \citealt{johnstone07}). Nevertheless, this discrepancy between the observed and the predicted \ha+\niiopab\, suggests caveats in our models, which could be related to the a priori assumptions.

Table~\ref{tab:listline} presents the list of the twenty brightest predicted line emission at different \av\, for the seven models described in this paper, ordered from the brightest to the fainter. The total energy released obviously depends on the parameters of a given model but is enough to power all the different lines. 

There is a range of specific models for which (i) the input X-ray energy is similar to the value estimated in \cite{sanders07} if coming from self-irradiation by the cooling gas, (ii) this energy is enough to power the total line emission, and produce outwards X-ray surface brightness comparable to the observed value for \av\,$\geq$5 mag. 

Finally, the actual models produce very faint or even no O\,{\sc vi}\,$\lambda$1032,1038$\AA$ doublet. For example, for \gx=100 the surface brightness of these lines is of the order of 10$^{-24}$ \ergcms\ arcsec$^{-2}$, while that of \ha\ is $\sim$10$^{-15}$ \ergcms\ arcsec$^{-2}$. This is in agreement with the no-detection of these lines shown by \cite{lecavelier04}. However, observations reveal a tension in this regard, since other works have detected O\,{\sc vi} doublet at the center of some cool-core clusters (e.g. \citealt{bregman06}). The O\,{\sc vi} line emission can be enhanced in radiative shocks and mixing layers (e.g. \citealt{mcquinn18}; \citealt{ji19}), which are not considered in our models.

\section{Conclusions}\label{sec:conclusions}
This article presents the modelling of the physical, chemical structure and line emission of filaments surrounding BCGs. We use the \cloudy\, code to revisit a self-irradiated model where the source of photoionization and heating of the filaments is the cooling radiation from the hot ICM plasma (the so-called cooling flow). The energy source is assumed to be the radiation from the gas cooling out of the hot plasma. We explore the effect of the penetration of the photons inside the clouds by varying \av. As a result, the grid of models predicts the emission lines arising from the ionized, neutral and molecular regions. We also explore turning on additional (turbulent) heating and we discuss the impact of such another source of heating that acts as an extra-pressure.

The structure of the clouds changes as the free parameter change. Increasing the intensity of the radiation field, the temperature and the ionization fraction of the gas also increase, while the density decreases (the models are isobaric). For low X-ray intensity the edge of the cloud is atomic and becomes molecular inside. Increasing the intensity, the ionization and molecular fronts move to higher \av. Increasing the turbulence has similar effects, while increasing the metallicity moves to lower \av\ the transition between the ionized and the atomic region. Changing the metallicity has less dramatic consequences on the model predictions. 

We note that the important parameters are (i) the slope of the input radiation field: the power law used in $\cloudy$ to model X-ray cooling flows that produces the soft X-ray and EUV photons that play a major role in the gas ionization; (ii) the strength of the input X-ray radiation field; (iii) the Av; and (iv) the level of turbulence, which provides extra-heating through dissipation of energy. There is an unavoidable degeneracy in the impact of these parameters, but it is clear that without any X-ray source it is not possible to reproduce the observations.

We have compared our grid of models to the rich range of observations now achieved in cool core cluster. Filaments in BCGs have been mapped from the soft X-ray to the millimeter and many lines in the optical and the  infrared have been detected. We have used such data to restrict the plausible range of parameter. 
The different combination of the free parameters presented in this paper can reproduce the multi-wavelength observables without requiring an excess of X-ray luminosity. The emitted intensity in the X-ray from \cite{sanders05} and \cite{walker15} is in agreement with the estimated reprocessed emitted intensity in the 2-10 keV band as well as 0.6-2 keV band of the models with \gx$\sim$10 and the X-ray input energy is enough to power the line emission.

Within a reduced range of parameter, some models can simultaneously reproduce most of the ionized and molecular lines. The combination of \gx\, between 1 to 10$^{3}$, any metallicity and turbulence that varies according to the selected \gx, for Av between 0.1 and 10 mag reproduces the typical BCG LINER-like low-ionization line ratio observed in BPT-diagrams, including [O\,{\sc iii}], [O\,{\sc i}] being more difficult to reproduce if not at very low Av (less then 0.1 mag). This restricted range of parameter space reproduces the infrared ratios as well.  
We note that the MIR and FIR lines arising from the atomic and molecular gas phases are very dependent on the \av. 
Finally, only the models with v$_{\rm tur}$=10 or 30 \kms\, and \gx$\geq$ 10 can reproduce the observed CO transitions.

It is clear that the mechanisms powering the nebula are more complex than these simple single component plane-parallel models of a constant pressure self-irradiated clouds. A better representation of the filaments may come from a combination of such models as expected for a population of clouds with a range of \av\ illuminated by different \gx\ and extra-heating. We are aware that a constant pressure model is a simplification and we also expect a contribution of energetic particles and of star formation in some regions but this was out of the scope of the actual study. In future, new constraints from high resolution multi-wavelength line emission as well as molecular absorption line, from Multi Unit Spectroscopic Explorer (MUSE) and Atacama Large Millimeter/submillimeter Array (ALMA), for several BCGs will help to developed a more complex modeling.

In a future paper we will use the restricted grid of model identified in this study to model the multi-wavelengths observations of the filamentary regions surrounding NGC\,1275. 

\begin{acknowledgements}
The authors thank the anonymous referee for providing helpful comments that contribute to improving the paper.
This work was supported by the ANR grant LYRICS (ANR-16-CE31-0011). The grids of models have been run on the computing cluster {\it Totoro} funded by the European Research Council, under the European Community’s Seventh framework Programme, through the Advanced Grant MIST (FP7/2017-2022, No 742719) 
\end{acknowledgements}

%
\bibliographystyle{aa} 
\bibliography{biblio} 
%

\appendix
\section{Chemical ISM composition}
The gas and dust elemental abundances used in our models are those given by \cloudy\, for the interstellar medium case. We report in Table~\ref{tab:abundances} the values for the gas-phase and in Table~\ref{tab:grains} the values for the dust conposition.
\begin{table}[htbp!]
\centering
     \caption{Gas-phase abundances used in our models.}
       \begin{tabular}{ l r l r}
       		\hline
       		\hline
        		 \noalign{\smallskip}
       		Element & Abundances & Element & Abundances\\
		\noalign{\smallskip}
			\hline
			\noalign{\smallskip}
			 He/H & 9.50$\times10^{-02}$  & Li/H & 5.40$\times10^{-11}$\\
			 Be/H & 1.00$\times10^{-20}$ & B/H & 8.90$\times10^{-11}$ \\
			 C/H & 3.00$\times10^{-04}$ & N/H & 7.00$\times10^{-05}$ \\
			 O/H & 4.00$\times10^{-04}$ & F/H & 1.00$\times10^{-20}$ \\
			 Ne/H & 6.00$\times10^{-05}$ & Na/H & 3.00$\times10^{-07}$ \\
			 Mg/H & 3.00$\times10^{-06}$ & Al/H & 2.00$\times10^{-07}$\\
			 Si/H & 4.00$\times10^{-06}$ & P/H & 1.60$\times10^{-07}$ \\
			 S/H & 1.00$\times10^{-05}$ & Cl/H & 1.00$\times10^{-07}$ \\
			 Ar/H & 3.00$\times10^{-06}$ & K/H & 1.10$\times10^{-08}$\\
			 Ca/H & 2.00$\times10^{-08}$ & Sc/H & 1.00$\times10^{-20}$ \\
			 Ti/H & 5.80$\times10^{-10}$ &  V/H & 1.00$\times10^{-10}$ \\
			 Cr/H & 1.00$\times10^{-08}$ & Mn/H & 2.30$\times10^{-08}$ \\
			 Fe/H & 3.00$\times10^{-06}$ & Co/H & 1.00$\times10^{-20}$ \\
			 Ni/H & 1.00$\times10^{-07}$ & Cu/H & 1.50$\times10^{-09}$ \\
			 Zn/H & 2.00$\times10^{-08}$ & & \\
	 	\hline
	 	\hline
	\end{tabular}
\\
 \label{tab:abundances}
\end{table}

\begin{table}[htbp!]
\centering
     \caption{Grain abundances used in our models.}
       \begin{tabular}{ l r l r}
       		\hline
       		\hline
        		 \noalign{\smallskip}
       		Element & Abundances & Element & Abundances\\
		\noalign{\smallskip}
			\hline
			\noalign{\smallskip}
			 C/H & 2.81$\times10^{-04}$ & O/H & 1.31$\times10^{-04}$ \\
			 Mg/H & 3.28$\times10^{-05}$ & Si/H & 3.28$\times10^{-05}$\\
			 Fe/H & 3.28$\times10^{-05}$ & &\\
	 	\hline
	 	\hline
	\end{tabular}
\\
 \label{tab:grains}
\end{table}

\section{Input parameters for the model with \gx=10, \vtur=10\kms\, and Z\,=1\,Z$_{\odot}$}
table HM05 z = 0\\
table SED "cool.sed"\\
intensity -1.8, range 14.71 to 147.06 Ryd\\
table sed "test$\_$modif.sed"\\
f(nu) -16.8611 0.1755\\
cosmic ray background 0\\
turbulence 10 km/s\\
hden 2\\
constant pressure no continuum set 6.5\\
no grain molecules\\
database H2\\
abundances he = -1.022 li = -10.268 be = -20.000 b = -10.051 c = -3.523 n = -4.155\\
continue o = -3.398 f = -20.000 ne = -4.222 na = -6.523 mg = -5.523 al = -6.699\\
continue si=-5.398 p=-6.796 s=-5.000 cl=-7.000 ar=-5.523 k = -7.959\\
continue ca = -7.699 sc = -20.000 ti = -9.237 v = -10.000 cr = -8.000 mn = -7.638\\
continue fe = -5.523 co = -20.000 ni = -7.000 cu = -8.824 zn = -7.6990 no grains\\
grains ism\\
grains pah\\
metals and grains 1 \\
set pah constant -4.6\\
set H2 Jura rate

\section{Comparison between our initial radiation field and the SED used by Ferland et al. 1994}
\begin{figure}[htbp!]
        \includegraphics[width=\hsize]{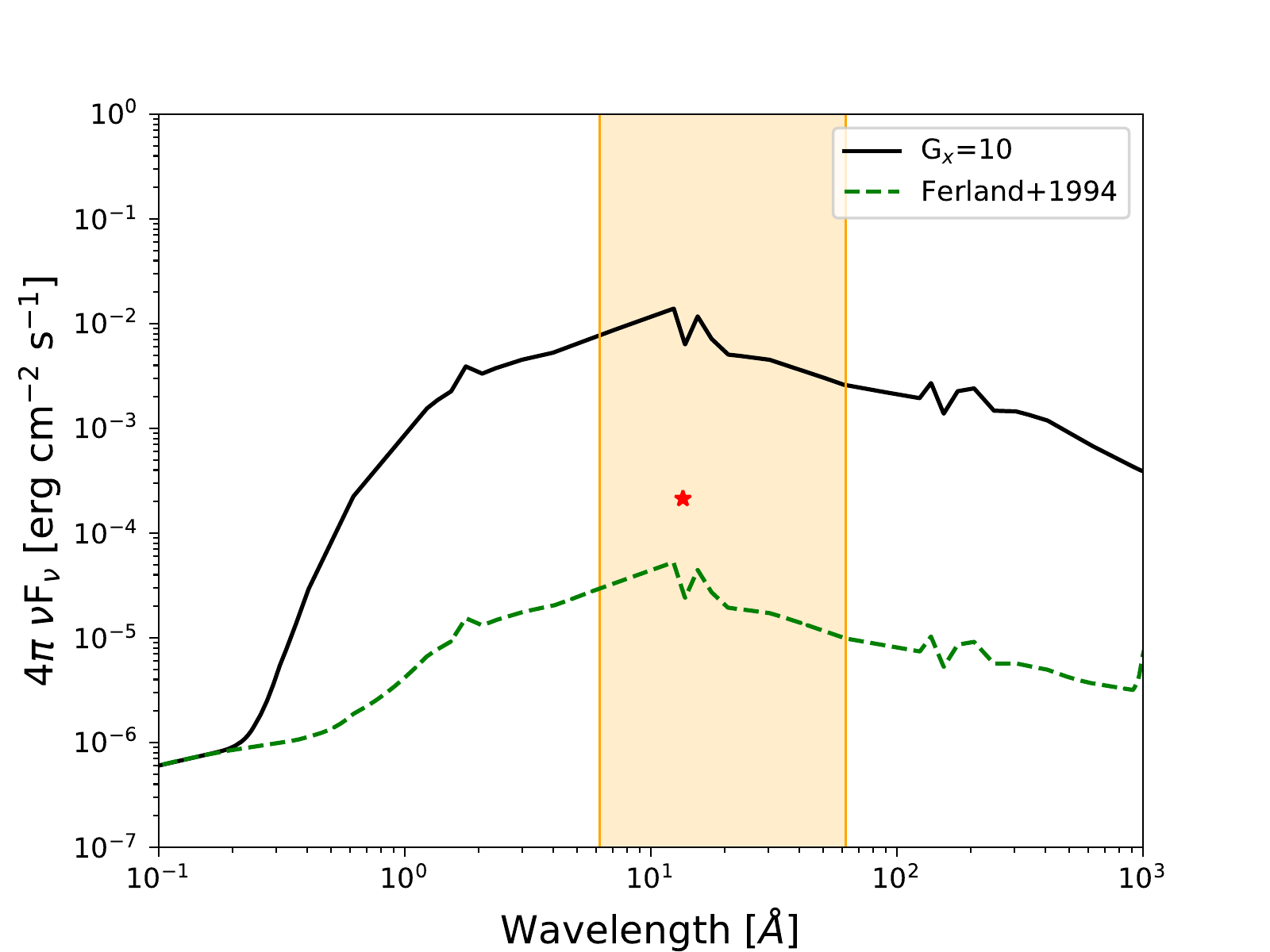}
        \caption{Comparison of the X-ray component of one of the input SED used in this study, i.e.\gx=10 (black curve), with that of the input SED used in \cite{ferland94} (green dashed curve). The red star represents the observed surface brightness measurements by \citealt{walker15}.}
    \label{fig:ferlandSED}
\end{figure}

Figure~\ref{fig:ferlandSED} shows the input SED used in \cite{ferland94}, in green, and the input SED of one of our models, i.e.\gx=10, as an example, in black. The input SED used by \cite{ferland94} is too faint compared to the more recent surface brightness measurements by \citealt{walker15} represented by the red star.

\section{Metallicity and turbulence effects in the structure of the cloud}
Figures~\ref{fig:exproplx2tur} and~\ref{fig:exproplx2tur} show the consequences of different  turbulent velocity on the structure of the cloud and Figures~\ref{fig:exproplx1met} and~\ref{fig:exproplx2met}  the effects of varying the metallicity. The Figures are described in Sec.~\ref{sec:physicsmodels}. 
\begin{figure*}[htbp!]
        \includegraphics[trim={2cm 6.5cm 0cm 7cm}, clip, width=\hsize]{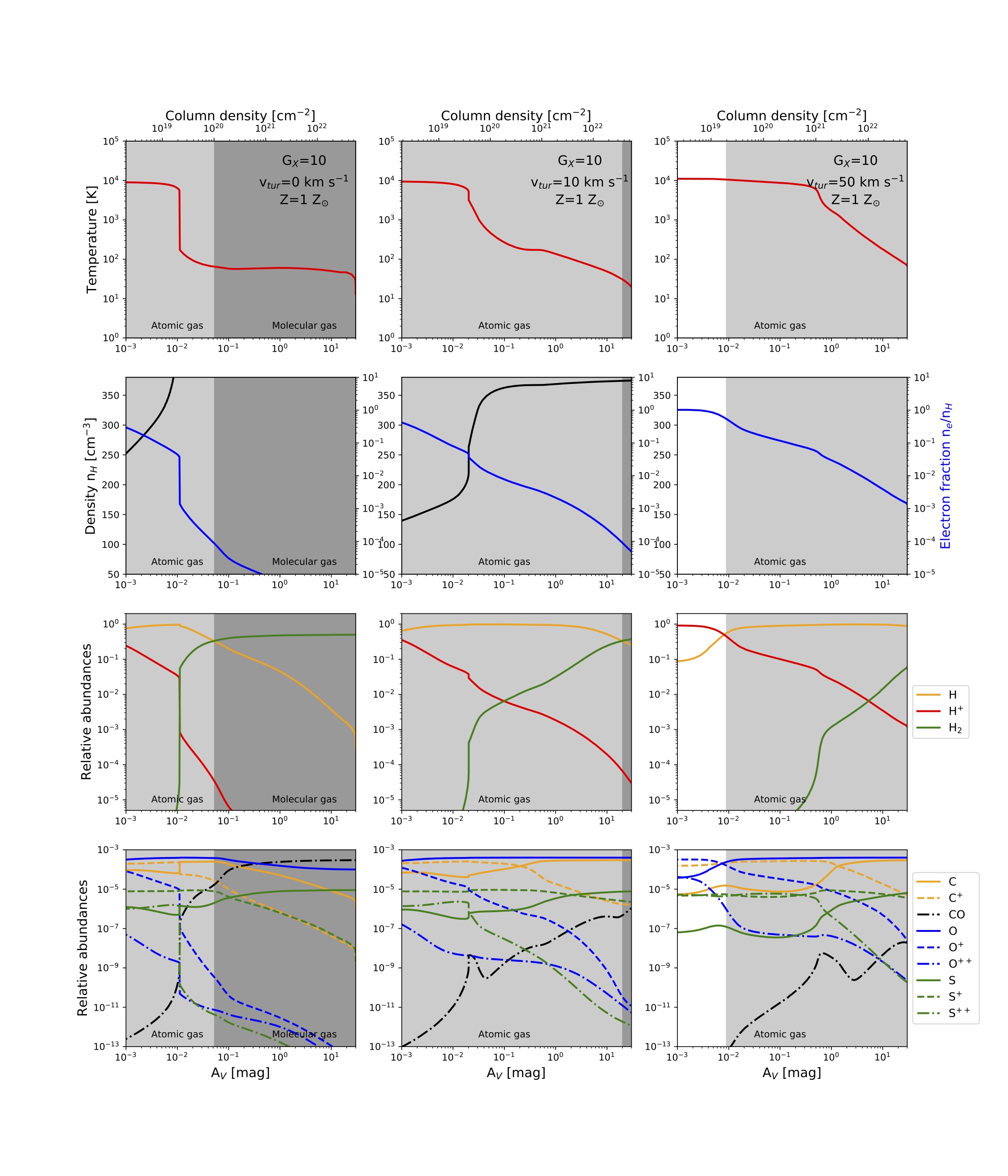}
        \caption{{\small Same as Fig.~\ref{fig:exproplx1}, but this time \gx is fixed at 10 and the turbulence varies: 0 km s$^{-1}$ (left column), 10 km s$^{-1}$ (central column) and 50 km s$^{-1}$ (right column).}}
    \label{fig:exproplx1tur}
\end{figure*}
\begin{figure*}[htbp!]
        \includegraphics[trim={2cm 5cm 0cm 5.5cm}, clip, width=\hsize]{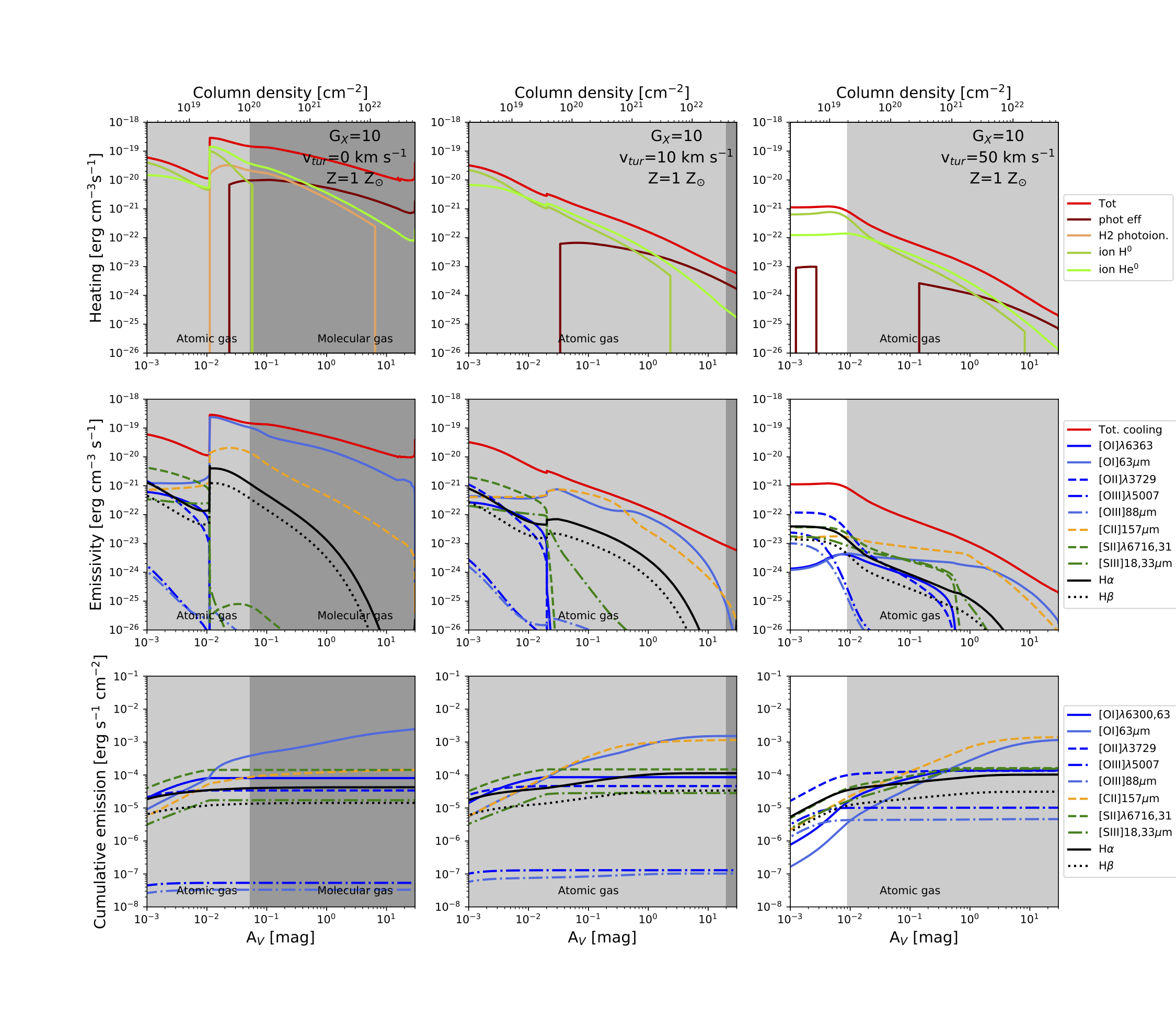}
        \caption{{\small Same as Fig.~\ref{fig:exproplx2}, but this time \gx is fixed at 10 and the turbulence varies: 0 km s$^{-1}$ (left column), 10 km s$^{-1}$ (central column) and 50 km s$^{-1}$ (right column).}}
    \label{fig:exproplx2tur}
\end{figure*}

\begin{figure*}[htbp!]
        \includegraphics[trim={2cm 6.5cm 0cm 7cm}, clip, width=\hsize]{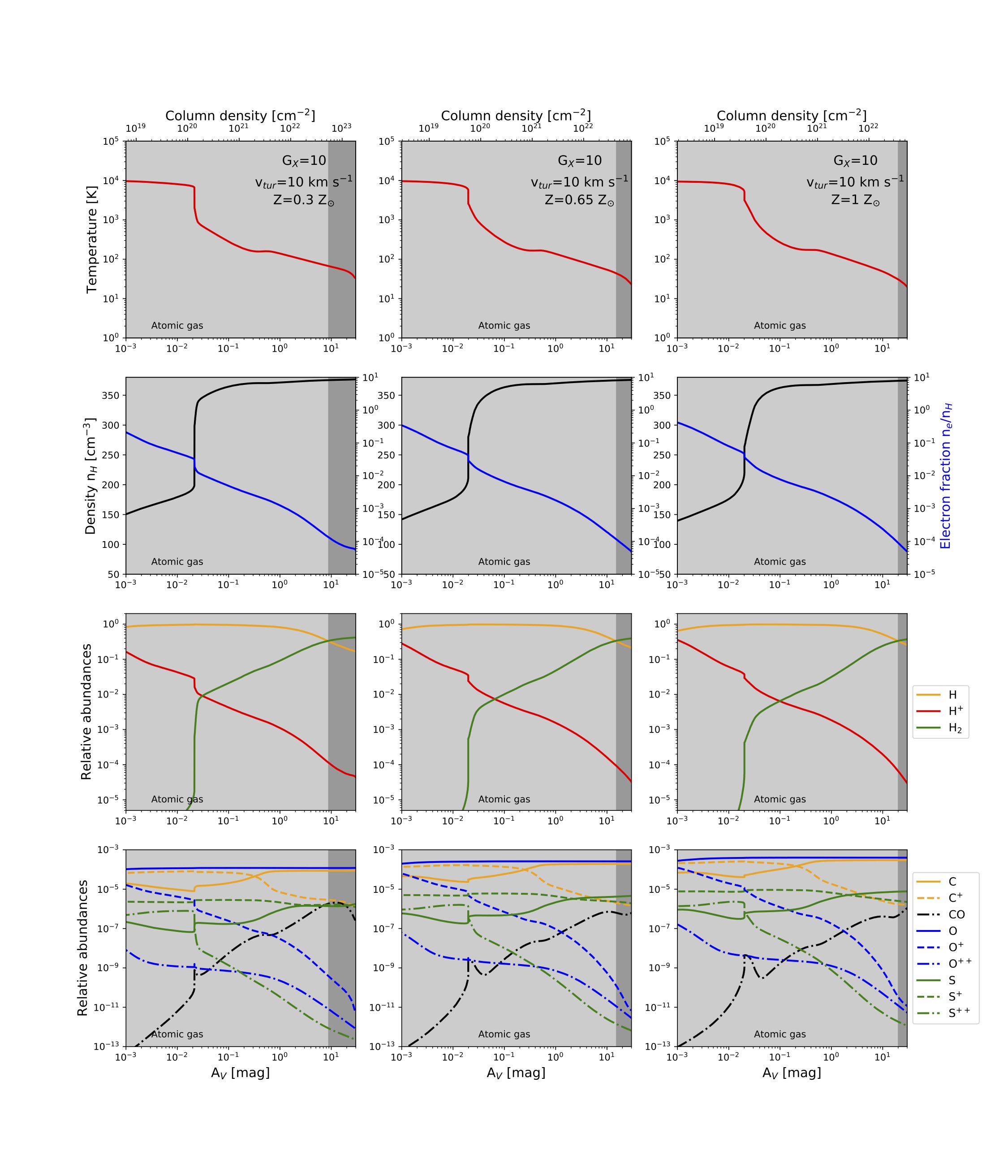}
        \caption{{\small Same as Fig.~\ref{fig:exproplx1}, but this time \gx is fixed at 10 and the metallicity varies: Z=0.3\,Z$_{\odot}$ (left column), Z=0.65\,Z$_{\odot}$ (central column) and Z=\,Z$_{\odot}$ (right column).}}
    \label{fig:exproplx1met}
\end{figure*}
\begin{figure*}[htbp!]
        \includegraphics[trim={2cm 5cm 0cm 5.5cm}, clip, width=\hsize]{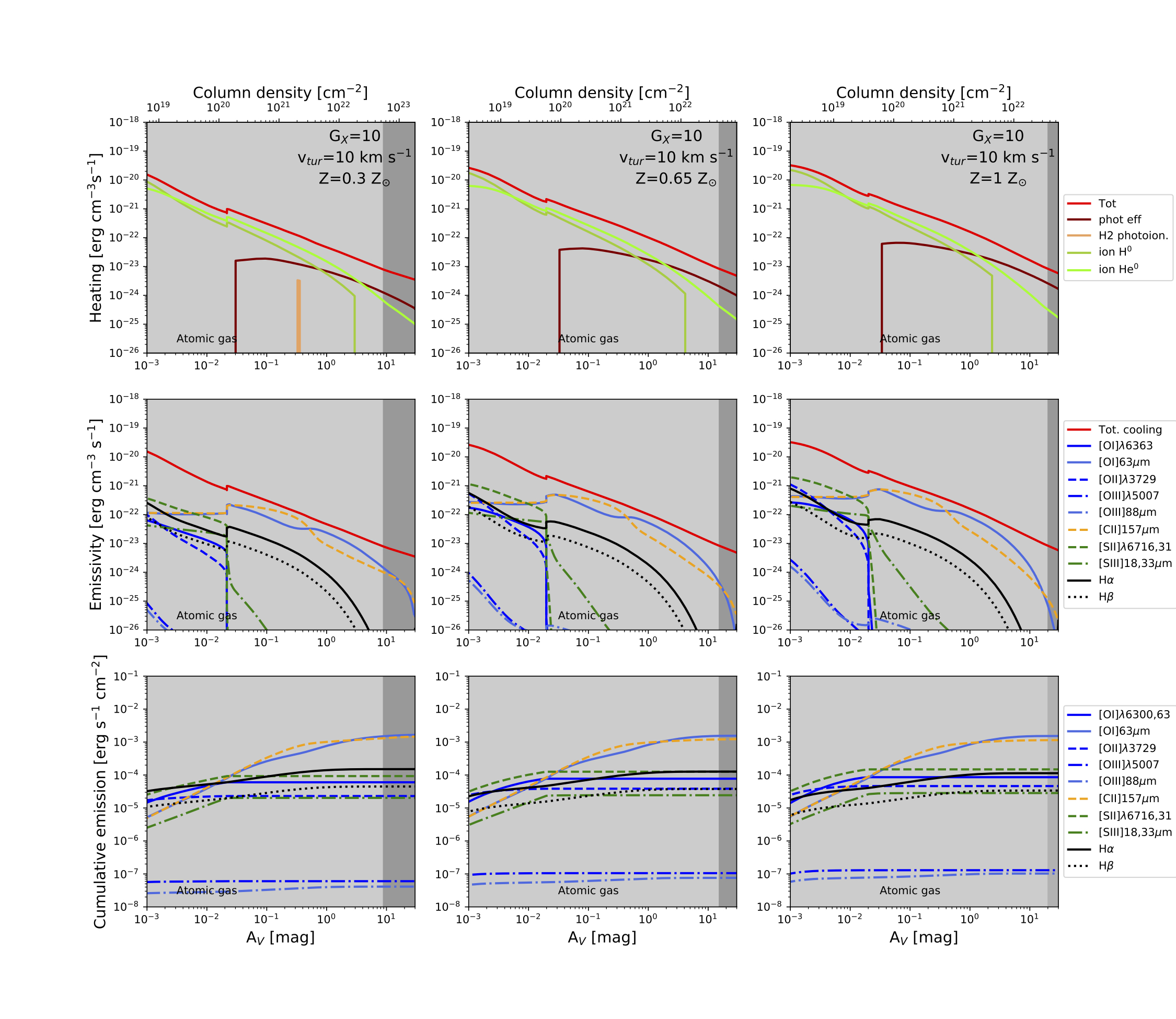}
        \caption{{\small Same as Fig.~\ref{fig:exproplx2}, but this time \gx is fixed at 10 and the metallicity varies: Z=0.3\,Z$_{\odot}$ (left column), Z=0.65\,Z$_{\odot}$ (central column) and Z=\,Z$_{\odot}$ (right column).}}
    \label{fig:exproplx2met}
\end{figure*}

\section{Behavior of \oiiiop/\hb, \niiop/\ha\ and [S\,{\sc ii}]/\ha\ in the parameter space}\label{sec:retiovsav}.
\begin{figure}[htbp!]
        \includegraphics[width=\hsize]{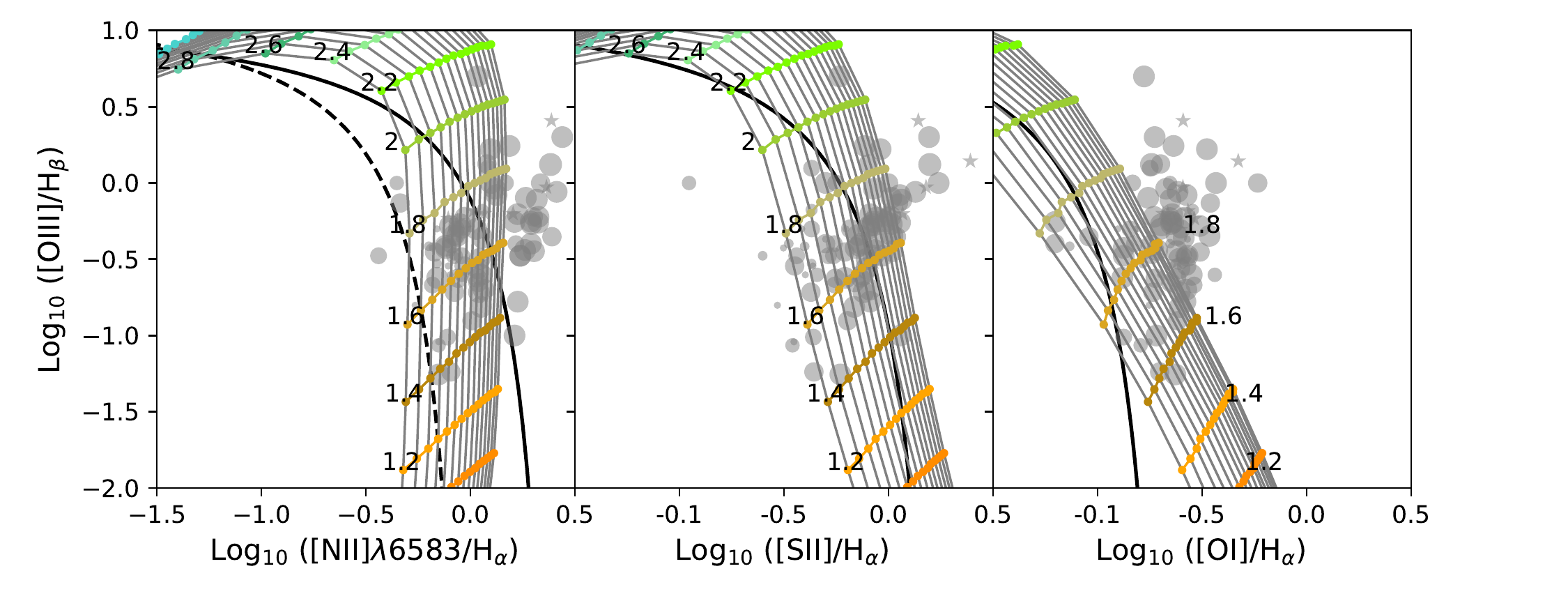}
        \includegraphics[width=\hsize]{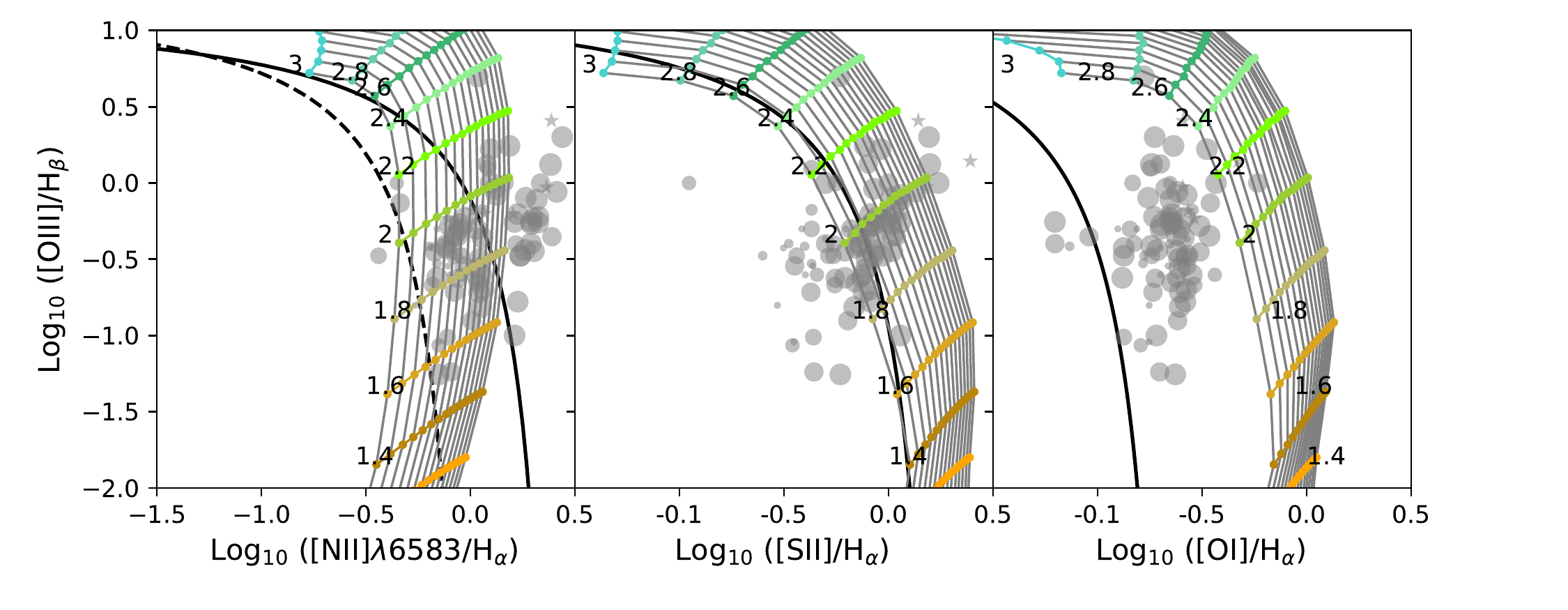}
        \includegraphics[width=\hsize]{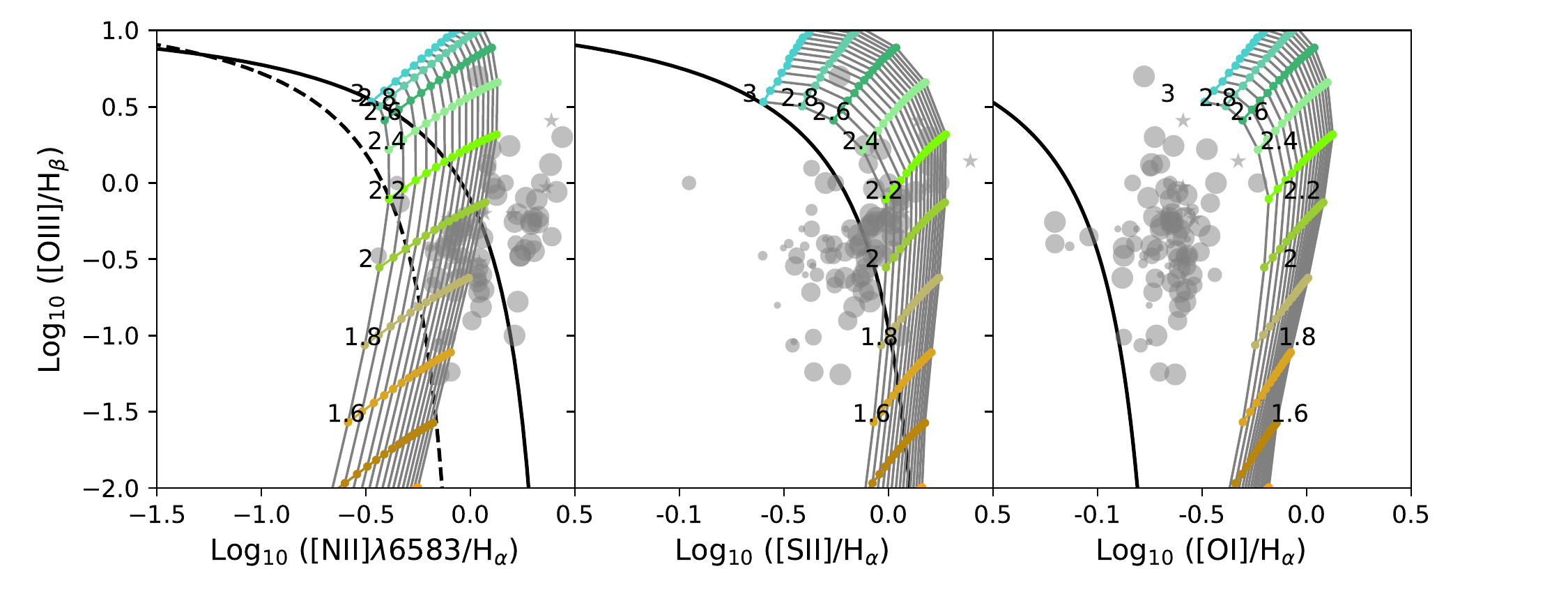}
        \includegraphics[width=\hsize]{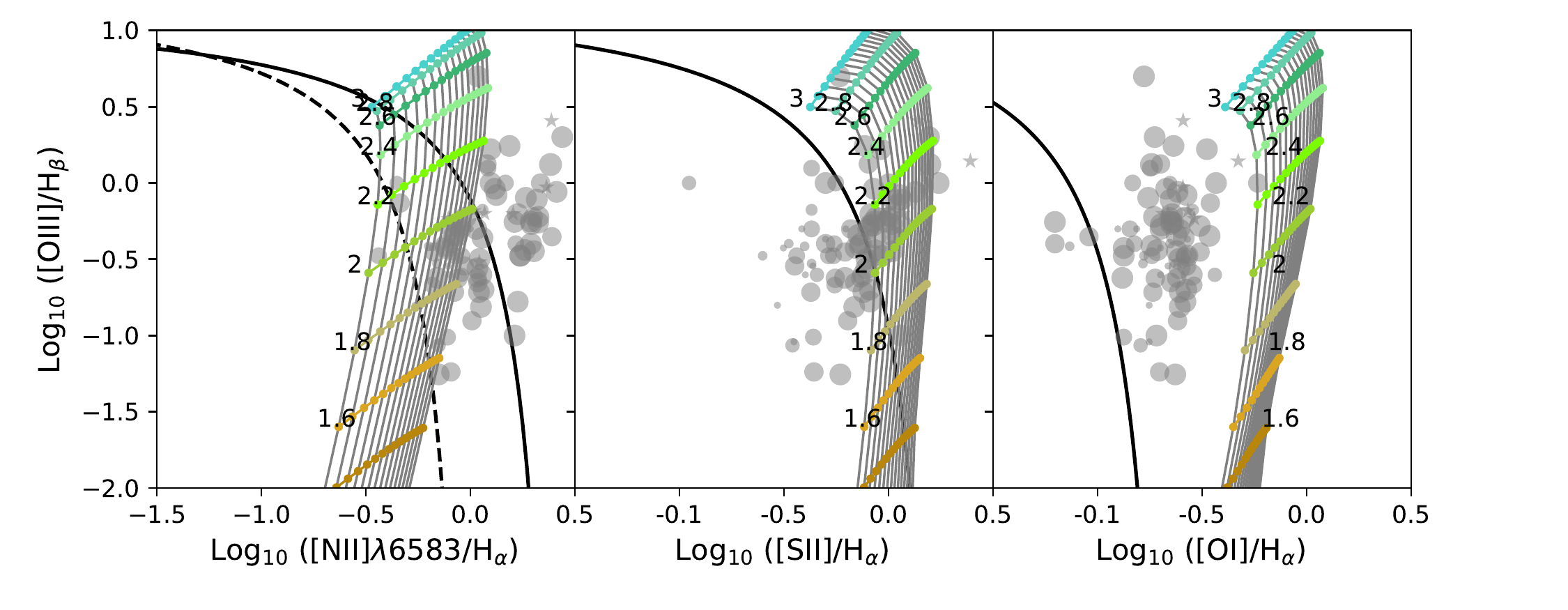}
        \caption{BPT diagrams. Comparison of the observed line ratios with the model predictions. The gray dots (filament values) and the stars (nucleus values) represent the data from \cite{mcdonald12}. The size of the dots decreases with the increase of the distance from the nuclei of the corresponding object. The solid black line is the upper limit for \hii\, regions by \cite{kewley01} and the dashed black line is the lower limit for AGN by \cite{kauffmann03}. The predicted cumulative line ratios from the models are overlaid. The gray lines correspond to a single value of metallicity and different values of X-ray emission (the logarithm of the value is written on the left and reproduced with different colours); from left to right the X-ray emission is constant and the metallicity increases (from 0.3 to 1 Z$_{\odot}$). The  turbulent velocity is fixed to 10 \kms. {\it Left column}: cumulative emission of \oiiiop/\hb\ vs. \niiop/\ha; {\it central column}: \oiiiop/\hb\ vs. [S\,{\sc ii}]/\ha; {\it right column}: \oiiiop/\hb\ vs. [O\,{\sc i}]/\ha. {\it Top row}: cumulative emission at \av\ = 0.001 mag; {\it second row}: cumulative emission at \av\ = 0.1 mag; {\it third row}: cumulative emission at \av\ = 1 mag; {\it Bottom row}:  cumulative emission at \av\ = 5 mag.}        
    \label{fig:bpt_1}
\end{figure}

\begin{figure}[htbp!]
        \includegraphics[width=\hsize]{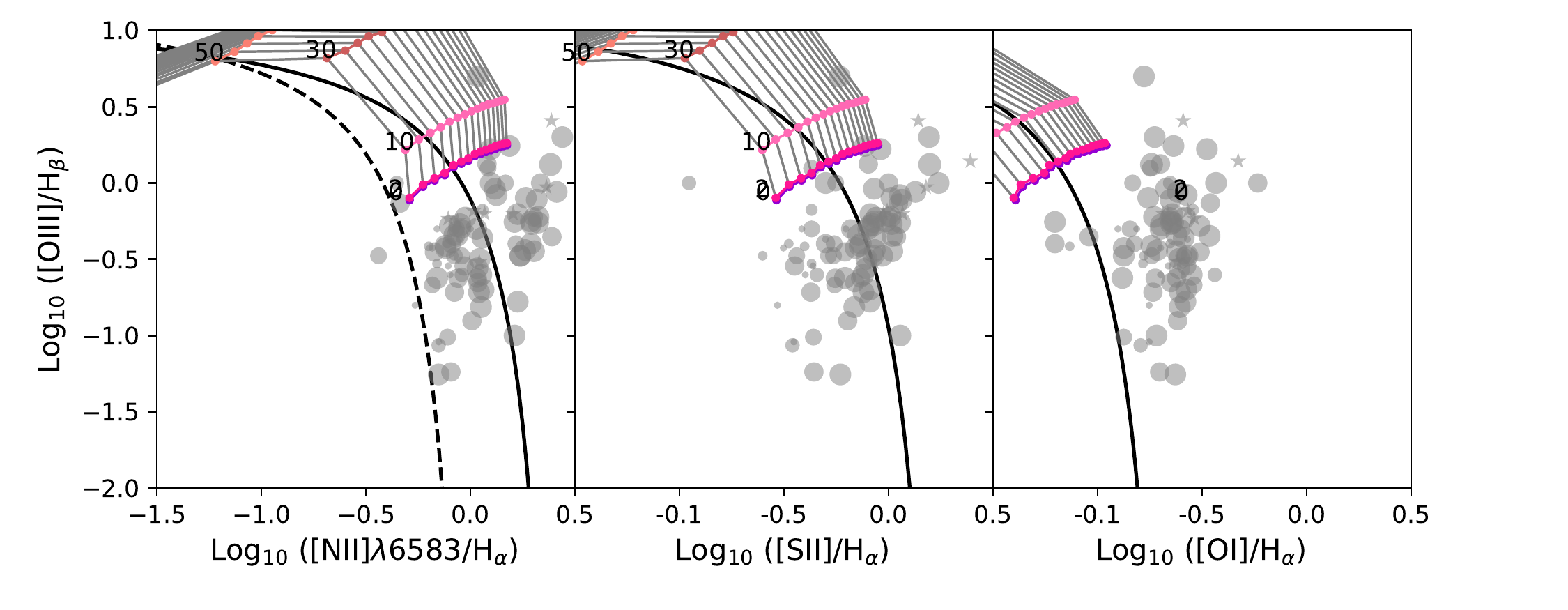}
        \includegraphics[width=\hsize]{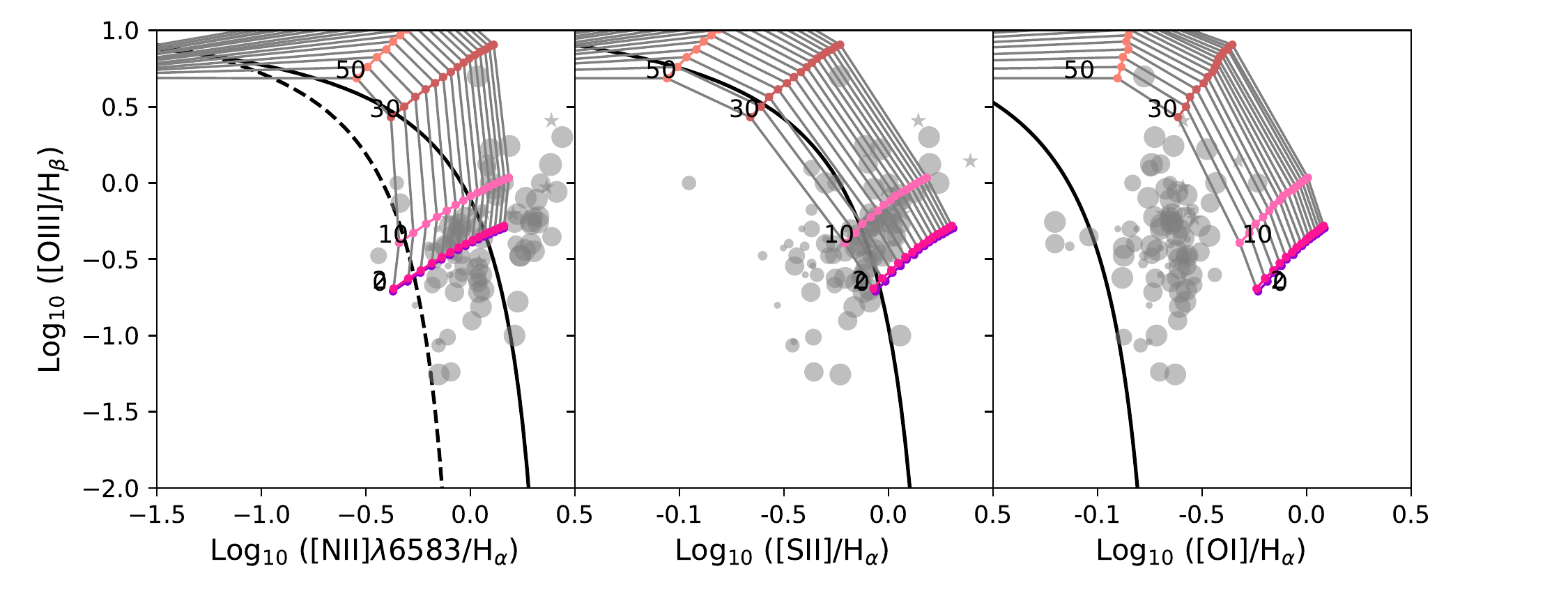}
        \includegraphics[width=\hsize]{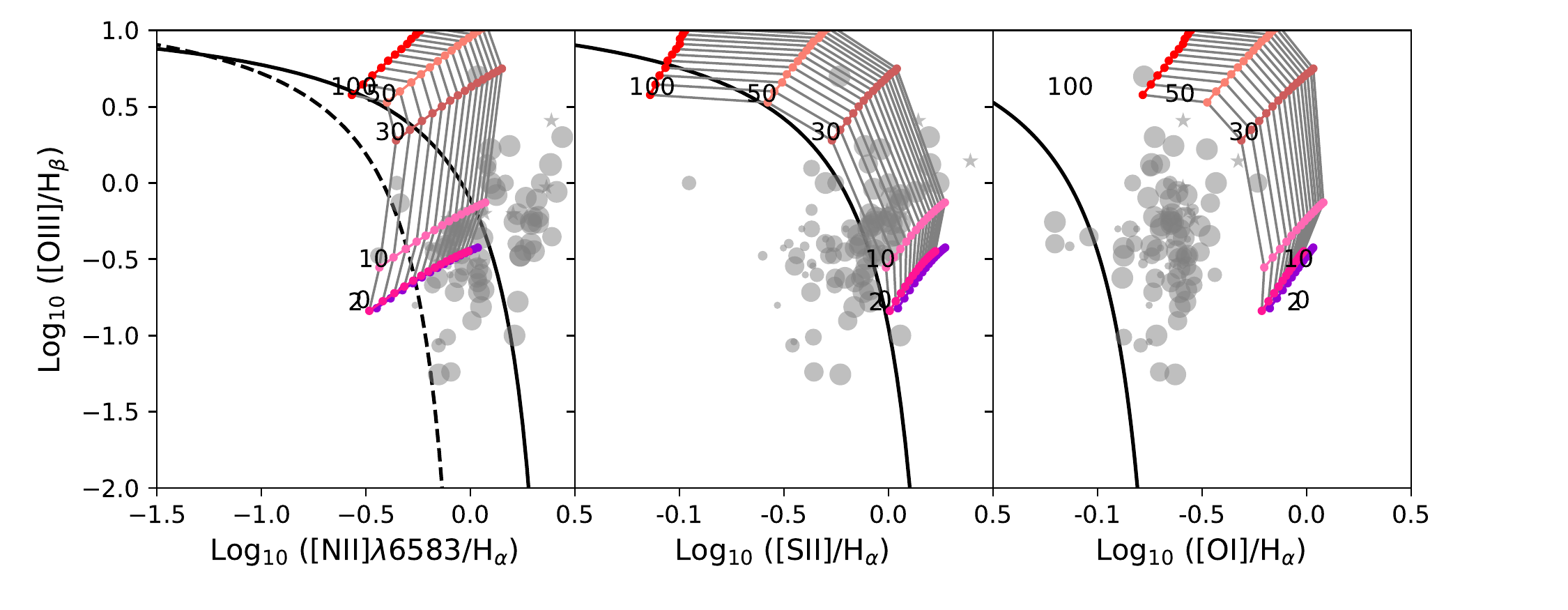}
        \includegraphics[width=\hsize]{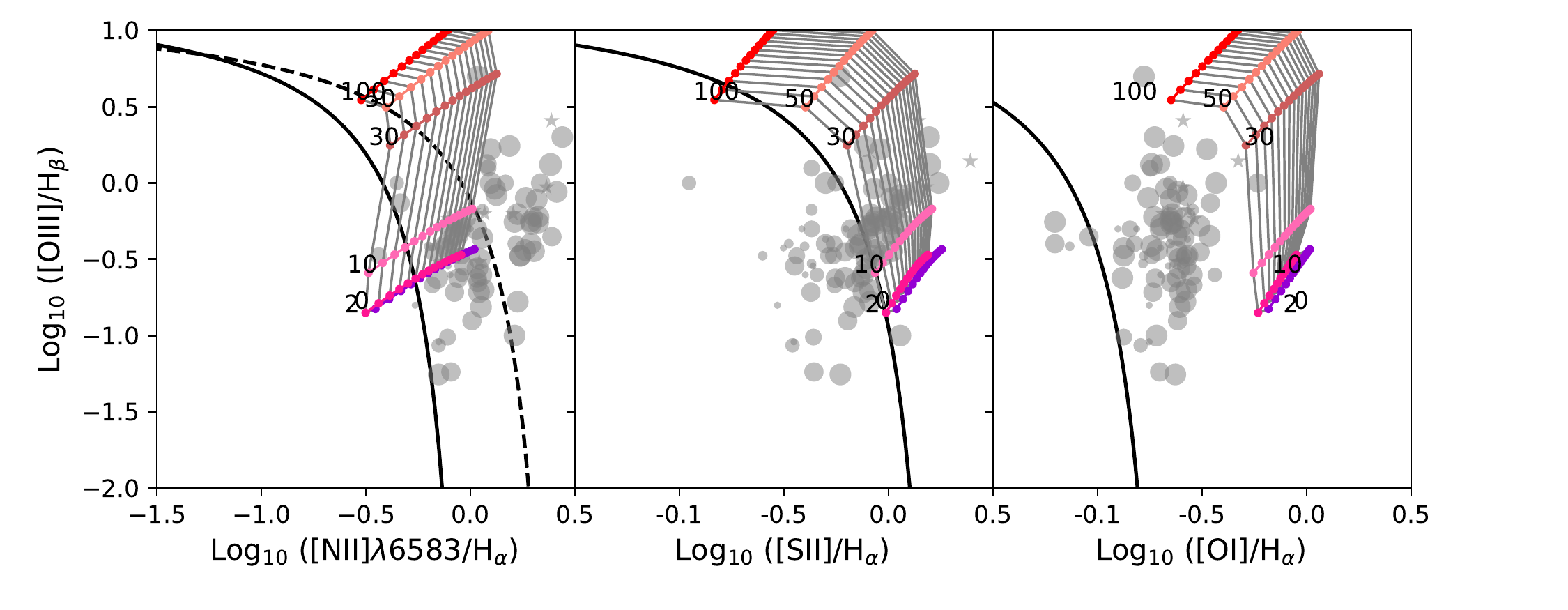}
        \caption{BPT diagrams. Comparison of the observed line ratios with the model predictions. The gray dots (filament values) and the stars (nucleus values) represent the data from \cite{mcdonald12}. The size of the dots decreases with the increase of the distance from the nuclei of the corresponding object. The solid black line is the upper limit for \hii\, regions by \cite{kewley01} and the dashed black line is the lower limit for AGN by \cite{kauffmann03}. The predicted cumulative line ratios from the models are overlaid. The gray lines correspond to a single value of metallicity and different values of turbulence (the velocity value is written on the left and reproduced with different colours); from left to right the velocity is constant and the metallicity increases (from 0.3 to 1 Z$_{\odot}$). The \gx\, is fixed to 100.{\it Left column}: cumulative emission of \oiiiop/\hb\ vs. \niiop/\ha; {\it central column}: \oiiiop/\hb\ vs. [S\,{\sc ii}]/\ha; {\it right column}: \oiiiop/\hb\ vs. [O\,{\sc i}]/\ha. {\it Top row}: cumulative emission at \av\ = 0.001 mag; {\it second row}: cumulative emission at \av\ = 0.1 mag; {\it third row}: cumulative emission at \av\ = 1 mag; {\it Bottom row}: cumulative emission at \av\ = 5 mag.}        
    \label{fig:bpt_2}
\end{figure}

\section{Predicted {\bf cumulative} line emission}
Table~\ref{tab:listline} presents the thirty brightest predicted line emission, from brightest to faintest, for the seven models described in the paper. The predicted line are for \av=0.1, 1, 5 and 30 mag.

\clearpage
\begin{table*}[htbp!]
\centering
     \caption{Brightest predicted {\bf cumulative} line emission.}
     \setlength{\tabcolsep}{4pt}
       {\small \begin{tabular}{ l l l | l c | l c | l c | l c}
       		\hline
       		\hline
        		 \noalign{\smallskip}
       		 \multicolumn{3}{l |}{Model} & \multicolumn{2}{c |}{\av = 0.1} & \multicolumn{2}{c |}{\av = 1} & \multicolumn{2}{c |}{\av = 5} & \multicolumn{2}{c}{\av = 30}\\
		\noalign{\smallskip}
			\hline
			 \gx & v$_{\rm tur}$ & Z & line & emission & line & emission & line & emission & line & emission\\
			  & km/s & Z$_{\odot}$ & & erg s$^{-1}$ cm$^{-2}$ & & erg s$^{-1}$ cm$^{-2}$ & & erg s$^{-1}$ cm$^{-2}$ & & erg s$^{-1}$ cm$^{-2}$\\
			\hline
			\noalign{\smallskip}
		 0.1 & 10 & 1 & [OI] 63$\mu$m & 7.13e-06 & [CI]  609$\mu$m & 1.31e-05 & [CI] 609$\mu$m & 3.36e-05 & [CI] 609$\mu$m & 9.29e-05 \\
		  
		  & & & \cii &  4.06e-06  & [OI] 63$\mu$m & 7.21e-06 & H2 6.91$\mu$m & 1.53e-05 & [CI] 370$\mu$m & 3.56e-05 \\   
		  
		  & & & [CI]  609$\mu$m & 3.24e-06 & H2 6.91$\mu$m & 5.65e-06 & H2 5.5$\mu$m & 1.37e-05 & H2 6.9$\mu$m & 2.51e-05 \\  
		  
		  & & & [CI] 370$\mu$m &  3.17e-06 & H2 5.51$\mu$m & 5.12e-06 & H2 4.69$\mu$m & 1.04e-05 & H2 5.5$\mu$m & 2.07e-05 \\
		  
		  & & & [SiII]  35$\mu$m &  1.40e-06  & [CI] 370$\mu$m & 5.00e-06 & H2 9.66$\mu$m & 1.01e-05 & H2 4.7$\mu$m & 1.45e-05 \\  
		  
		  & & & [NaI] 5890$\AA$ & 1.05e-06 & H2 9.66$\mu$m & 4.68e-06 & [CI] 370$\mu$m &  8.61e-06 & CO 1300.05$\mu$m & 1.18e-05 \\  
		  
		  & & & H2 9.66$\mu$m & 9.84e-07  & \cii & 4.23e-06 & [OI] 63$\mu$m &  7.36e-06 & H2 9.66$\mu$m &  1.06e-05 \\  
		  
		  & & & [NaI] 5896$\AA$ & 9.31e-07 & H2 4.69$\mu$m & 3.96e-06 & H2 17$\mu$m & 7.21e-06 & H2 17$\mu$m &  1.05e-05 \\
		  
		  & & & H2 6.91$\mu$m & 7.73e-07 & H2 17$\mu$m & 2.94e-06 & H2 4.17$\mu$m & 6.89e-06 & H2 4.17$\mu$m & 8.75e-06 \\  
		  
		  & & & H2 17$\mu$m & 7.11e-07 & H2 4.17$\mu$m & 2.68e-06 & H2 1.74$\mu$m & 5.24e-06 & [OI] 63$\mu$m & 7.65e-06 \\  
		  
		  & & & H2 5.51$\mu$m & 5.89e-07  & H2 1.74$\mu$m &  2.48e-06 & H2 6.1$\mu$m & 4.63e-06 & H2 6.1$\mu$m & 7.52e-06 \\  
		  
		  & & & HI 6562.81$\AA$ & 5.43e-07 & H2 3.84$\mu$m & 1.80e-06 & H2 3.84$\mu$m & 4.53e-06 & H2 8.02$\mu$m & 6.46e-06 \\
		  
		  & & & [OI] 145$\mu$m & 5.13e-07 & H2 1.83$\mu$m &  1.77e-06  & H2 4.4$\mu$m & 4.45e-06 & CO 866.727$\mu$m & 6.35e-06 \\  
		  
		  & & & H2 4.69$\mu$m & 4.30e-07  & H2 4.4$\mu$m & 1.72e-06 & \cii & 4.38e-06 & H2 4.4$\mu$m & 5.78e-06 \\  
		  
		  & & & H2 12$\mu$m & 3.28e-07  & H2 6.1$\mu$m &  1.71e-06 & H2 8.0$\mu$m & 4.34e-06 & H2 5.05$\mu$m & 5.64e-06 \\  
		  
		  & & & H2 1.74$\mu$m & 3.17e-07 &  H2 8.02$\mu$m & 1.70e-06 & H2 4.07$\mu$m & 4.19e-06 & H2 1.74$\mu$m &  5.37e-06 \\
		  
		  & & & [FeII] 26$\mu$m & 3.07e-07 & H2 4.07$\mu$m &  1.67e-06  & H2 1.83$\mu$m & 3.88e-06 & H2 3.84$\mu$m & 5.32e-06 \\   
		  
		  & & & H2 8.02$\mu$m & 2.92e-07  & H2 3.83$\mu$m & 1.57e-06  & H2 3.83$\mu$m & 3.86e-06 & H2 4.95$\mu$m & 5.19e-06 \\   
		  
		  & & & [MgI] 2852$\AA$ & 2.86e-07 & H2 1.12$\mu$m & 1.48e-06 & H2 5.05$\mu$m &  3.83e-06 & H2 4.07$\mu$m & 4.85e-06 \\   
		  
		  & & & H2 4.17$\mu$m & 2.84e-07 & H2 2.56$\mu$m & 1.47e-06 & H2 4.9$\mu$m & 3.55e-06 & H2 12$\mu$m & 4.83e-06 \\

		  \noalign{\smallskip}
		 \hline
		 \noalign{\smallskip}
		 10 & 10 & 1 & \cii & 3.85e-04  & \cii & 9.77e-04 & [OI] 63$\mu$m & 1.29e-03 & [OI] 63$\mu$m & 1.52e-03 \\  
		  
		  & & & [OI] 63$\mu$m & 2.86e-04 & [OI] 63$\mu$m & 8.22e-04 & \cii & 1.11e-03 & H2 6.9$\mu$m & 1.31e-03 \\  
		  
		  & & & [SII] 6716$\AA$ & 1.15e-04 & [SiII] 35$\mu$m & 1.87e-04 & [CI] 370$\mu$m & 5.67e-04 & [CI] 370$\mu$m & 1.28e-03 \\  
		  
		  & & & [CI] 9850$\AA$ & 1.14e-04 & [CI] 370$\mu$m & 1.40e-04 & H2 6.9$\mu$m & 2.55e-04 & H2 5.5$\mu$m & 1.17e-03\\  
		  
		  & & & [SiII] 35$\mu$m & 1.08e-04  & HI 6563$\AA$ & 1.26e-04 & [CI] 609$\mu$m & 2.54e-04 & \cii &  1.15e-03\\   
		  
		  & & & [OI] 6300$\AA$ &  8.69e-05  & [CI] 9850$\AA$ & 9.79e-05 & H2 5.5$\mu$m & 2.39e-04 & H2 4.7$\mu$m & 8.77e-04\\ 
		  
		  & & & HI 6563$\AA$ &  8.53e-05  &  [SII] 6716$\AA$ & 9.64e-05 & [SiII] 35$\mu$m & 2.01e-04 & [CI] 609$\mu$m & 5.85e-04 \\  
		  
		  & & & [SII] 6731$\AA$ & 8.24e-05 & [OI] 6300$\AA$ & 7.21e-05 & H2 4.69$\mu$m & 1.87e-04 & H2 4.17$\mu$m & 5.81e-04 \\  
		  
		  & & & [NII] 6583$\AA$ & 6.52e-05 & [SII] 6731$\AA$ & 6.89e-05 & H2 9.66$\mu$m & 1.57e-04 & H2 9.66$\mu$m & 4.27e-04 \\  
		  
		  & & & [NeII] 13$\mu$m &  6.25e-05 & [NeII] 13$\mu$m & 6.15e-05 & H2 4.17$\mu$m & 1.31e-04 & H2 4.06$\mu$m & 4.17e-04 \\  
		  
		  & & & [OII] 3729$\AA$ &  6.07e-05 & [OI] 145$\mu$m & 5.62e-05 & HI 6563$\AA$ & 1.20e-04 & H2 17$\mu$m & 4.14e-04\\  
		  
		  & & & [FeII] 5.3$\mu$m &  4.70e-05 & [NII] 6583$\AA$ &  5.44e-05 & H2 3.83$\mu$m & 9.48e-05 & H2 3.83$\mu$m & 4.09e-04\\ 
		  
		  & & & [OII] 3726$\AA$ &  4.27e-05 & [CI] 609$\mu$m & 5.24e-05 & H2 4.07$\mu$m & 9.37e-05 & H2 4.4$\mu$m & 4.01e-04\\   
		  
		  & & & [CI] 9824$\AA$ &  3.81e-05 & [OII] 3729$\AA$ & 4.92e-05 & H2 3.84$\mu$m & 9.12e-05 & H2 3.84$\mu$m & 3.84e-04 \\  
		  
		  & & & [NI] 5200$\AA$ &  3.52e-05 & [FeII] 5.3$\mu$m & 4.59e-05 & [SII] 6716$\AA$ & 8.73e-05 & H2 6.1$\mu$m & 3.60e-04 \\  
		  
		  & & & [FeII] 1.2$\mu$m & 3.36e-05 & HI 4861$\AA$ &  3.78e-05 & H2 4.4$\mu$m & 8.72e-05 & H2 5.05$\mu$m & 3.08e-04 \\  
		  
		  & & & [NeIII] 15$\mu$m & 2.94e-05 & [FeII] 25$\mu$m & 3.57e-05 & [CI] 9850$\AA$ & 8.63e-05 & H2 4.3$\mu$m & 3.05e-04\\   
		  
		  & & & [FeII] 1.6$\mu$m & 2.87e-05 & [OII] 3726$\AA$ &  3.47e-05 & H2 17$\mu$m & 8.16e-05 & H2 4.9$\mu$m & 2.95e-04\\ 
		  
		  & & & HI 4861$\AA$ & 2.80e-05 & [CI] 9824$\AA$ & 3.28e-05 &  H2 1.74$\mu$m & 8.03e-05 & H2 3.6$\mu$m & 2.72e-04\\  
		  
		  & & & [OI] 6363$\AA$ & 2.78e-05 & H2 6.9$\mu$m & 3.26e-05 & [OI] 145$\mu$m & 7.73e-05 & H2 8.02$\mu$m & 2.67e-04\\  
		  
	 	\hline
	 	\hline
	\end{tabular}}
\label{tab:listline}
\end{table*}

\begin{table*}[htbp!]
\centering
     \caption{Brightest {\bf cumulative} predicted line emission.}
     \setlength{\tabcolsep}{4pt}
       {\small \begin{tabular}{ l l l | l c | l c | l c | l c}
       		\hline
       		\hline
        		 \noalign{\smallskip}
       		 \multicolumn{3}{l |}{Model} & \multicolumn{2}{c |}{\av = 0.1} & \multicolumn{2}{c |}{\av = 1} & \multicolumn{2}{c |}{\av = 5} & \multicolumn{2}{c}{\av = 30}\\
		\noalign{\smallskip}
			\hline
			 \gx & v$_{\rm tur}$ & Z & line & emission & line & emission & line & emission & line & emission\\
			  & km/s & Z$_{\odot}$ & & erg s$^{-1}$ cm$^{-2}$ & & erg s$^{-1}$ cm$^{-2}$ & & erg s$^{-1}$ cm$^{-2}$ & & erg s$^{-1}$ cm$^{-2}$\\
			\hline
			\noalign{\smallskip}
		100 & 10 & 1 & [OII] 3729$\AA$ & 1.72e-03 & \cii & 4.41e-03 & [OI] 63$\mu$m & 1.26e-02 & [OI] 63$\mu$m & 1.92e-02\\
		
		& & & [OII] 3726$\AA$ & 1.24e-03 & [OI] 63$\mu$m & 4.23e-03 & \cii & 7.11e-03 & \cii & 7.98e-03\\  
		
		& & & [NII] 6583$\AA$ & 1.20e-03 & [SiII] 35$\mu$m & 1.56e-03 & [SiII] 35$\mu$m & 3.32e-03 & [SiII] 35$\mu$m & 3.18e-03\\  
		
		& & & [OI] 6300$\AA$ & 8.22e-04 & [OII] 3729$\AA$ & 1.46e-03 & [NeII] 13$\mu$m & 1.65e-03 & H2 6.9$\mu$m & 3.13e-03\\ 
		
		& & & HI 6563$\AA$ & 7.91e-04 & [NeII] 13$\mu$m & 1.30e-03 & [OII] 3729$\AA$ & 1.36e-03 & [CI] 370$\mu$m & 2.89e-03\\  
		
		& & & [SII] 6716$\AA$ & 7.05e-04 & [SIII] 33$\mu$m & 1.29e-03 & [SIII] 33$\mu$m & 1.23e-03 & H2 5.5$\mu$m & 2.65e-03\\
		
		& & & [SIII] 9530$\AA$ & 5.93e-04 & [OI] 6300$\AA$ & 1.23e-03 & HI 6563$\AA$ & 1.10e-03 & H2 4.69$\mu$m & 1.95e-03\\  
		
		& & & \cii & 5.53e-04 & [NII] 6583$\AA$ & 1.18e-03 & [OI] 6300$\AA$ & 1.07e-03 & [OI] 145$\mu$m & 1.48e-03\\  
		
		& & & [SIII] 33$\AA$ & 5.24e-04 &  [SII] 6716$\AA$ & 1.13e-03 & [OI] 145$\mu$m & 1.07e-03 & [OII] 3729$\AA$ & 1.36e-03\\    
		
		& & & [SII] 6731$\AA$ & 5.12e-04 & HI 6563$\AA$ & 1.08e-03 & [NII] 6583$\AA$ & 1.05e-03 & H2 4.17$\mu$m & 1.31e-03\\  
		
		& & & [OI] 63$\mu$m & 4.33e-04 & [OII] 3726$\AA$ & 1.05e-03 & [OII] 3726$\AA$ & 9.79e-04 & H2 9.66$\mu$m & 1.30e-03\\  
		
		& & & HeII 1640$\AA$ & 4.11e-04 & [FeII] 5.3$\mu$m & 9.33e-04 & [SII] 6716$\AA$ & 9.70e-04 & [NeII] 13$\mu$m & 1.25e-03\\
		
		& & & HeII 304$\AA$ &  4.11e-04 & [SIII] 9530$\AA$ & 9.04e-04 & [FeII] 5.3$\mu$m & 8.93e-04 & [OI] 6300$\AA$ & 1.06e-03\\  
		
		& & & [NII] 6548$\AA$ & 4.08e-04 & [SII] 6731$\AA$ & 8.02e-04 & [CI] 370$\mu$m & 7.94e-04 & [CI] 609$\mu$m & 1.06e-03\\  
		
		& & & [NeII] 15$\mu$m &  3.35e-04 & [SIII] 19$\mu$m & 7.09e-04 & [SIII] 9530$\AA$ & 7.69e-04 & [NII] 6583$\AA$ & 1.04e-03\\  
		
		& & & [CII] 2325$\AA$ &  3.30e-04 & [NeIII] 15$\mu$m & 6.85e-04 & [FeII] 25$\mu$m & 7.64e-04 & HI 6563$\AA$ & 1.02e-03\\ 
		
		& & & [SIII] 19$\mu$m & 3.14e-04 & [FeII] 1.2$\mu$m & 5.44e-04 & [SII] 6731$\AA$ & 6.90e-04 & H2 3.8$\mu$m & 1.01e-03\\  
		
		& & & [NI] 5200$\AA$ & 2.89e-04 & [FeII] 1.6$\mu$m & 4.83e-04 & [NeIII] 15$\mu$m & 6.90e-04 & H2 4.06$\mu$m & 1.00e-03\\
		
		& & & [FeII] 5.3$\mu$m & 2.86e-04 & [NI] 5200$\AA$ & 4.70e-04 & H2 9.66$\mu$m & 6.89e-04 & [SIII] 33$\mu$m & 1.00e-03\\  
		
		& & & [SiII] 35$\mu$m & 2.84e-04 & [ArII] 7.0$\mu$m & 4.67e-04 & [SIII] 19$\mu$m & 6.37e-04 & [OII] 3726$\AA$ & 9.79e-04\\  
		\noalign{\smallskip}
		\hline
		\noalign{\smallskip}
		10 & 0 & 1 & [OI] 63$\mu$m & 4.62e-04 & [OI] 63$\mu$m & 9.14e-04 & [OI] 63$\mu$m & 1.66e-03  & [OI] 63$\mu$m & 2.46e-03\\  
		
		& & & [CI] 9850$\AA$ & 1.49e-04 & H2 1.21$\mu$m & 4.79e-04 & H2 2.12$\mu$m & 1.08e-03 & H2 4.69$\mu$m & 1.49e-03\\  
		
		& & & [SII] 6716$\AA$ & 1.10e-04 & H2 4.69$\mu$m & 3.77e-04 & H2 4.69$\mu$m & 1.01e-03 & H2 2.12$\mu$m & 1.18e-03\\ 
		
		& & & \cii & 1.00e-04 & H2 2.42$\mu$m & 3.48e-04 & H2 3.48$\mu$m & 8.36e-04 & H2 4.17$\mu$m & 1.15e-03\\  
		
		& & & [OI] 6300$\AA$ & 8.12e-05 & H2 3.48$\mu$m & 3.20e-04 & H2 2.42$\mu$m & 8.28e-04 & H2 3.48$\mu$m & 1.13e-03\\  
		
		& & & [SII] 6731$\AA$ & 8.03e-05 & H2 4.17$\mu$m & 3.10e-04 & H2 4.18$\mu$m & 8.18e-04 & H2 2.42$\mu$m & 9.64e-04\\
		
		& & & HI 6563$\AA$ & 5.44e-05 & H2 2.40$\mu$m & 2.85e-04  & H2 2.40$\mu$m & 6.97e-04  & H2 3.6$\mu$m & 8.73e-04\\  
		
		& & & H2 2.12$\mu$m & 5.43e-05 & H2 1.34$\mu$m & 2.64e-04 &  H2 3.6$\mu$m & 6.53e-04  & H2 2.4$\mu$m & 8.11e-04\\  
		
		& & & [NII] 6583$\AA$ & 5.21e-05 & H2 3.6$\mu$m & 2.52e-04 & H2 2.8$\mu$m & 6.29e-04  & H2 2.8$\mu$m & 7.94e-04\\  
		
		& & & [CI] 9824$\AA$ & 5.00e-05 & H2 1.21$\mu$m & 2.47e-04 & H2 1.34$\mu$m & 5.56e-04  & H2 3.8$\mu$m & 7.11e-04\\  
		
		& & & [SiII] 35$\mu$m & 4.96e-05 & H2 2.8$\mu$m & 2.45e-04 & H2 3.84$\mu$m & 5.26e-04 & H2 3.2$\mu$m & 6.62e-04\\  
		
		& & & [OII] 3729$\AA$ & 4.47e-05 & H2 3.8$\mu$m & 2.03e-04 & H2 3.23$\mu$m & 5.08e-04 & H2 8.0$\mu$m & 6.11e-04\\
		
		& & & H2 4.69$\mu$m & 4.46e-05 & H2 3.2$\mu$m & 1.99e-04 & H2 1.21$\mu$m & 5.00e-04  & H2 4.4$\mu$m & 6.07e-04\\  
		
		& & & [NeII] 13$\mu$m & 4.17e-05 & H2 1.95$\mu$m & 1.80e-04 & H2 3.4$\mu$m & 4.35e-04  & H2 3.4$\mu$m & 5.89e-04\\  
		
		& & & H2 3.48$\mu$m & 3.98e-05 & H2 3.4$\mu$m & 1.72e-04 & H2 4.4$\mu$m & 4.14e-04 & H2 1.34$\mu$m & 5.51e-04\\  
		
		& & & [FeII] 5.3$\mu$m & 3.97e-05 & H2 9.66$\mu$m & 1.71e-04 & H2 1.95$\mu$m & 3.40e-04 & H2 6.1$\mu$m & 5.25e-04\\  
		
		& & & H2 2.42$\mu$m & 3.89e-05 & H2 2.24$\mu$m & 1.68e-04 & H2 2.24$\mu$m & 3.40e-04 & H2 6.9$\mu$m & 5.10e-04\\        
		
		& & & H2 4.17$\mu$m & 3.82e-05 & CO 325$\mu$m & 1.59e-04 & H2 9.66$\mu$m & 3.96e-04 & H2 2.24$\mu$m & 4.49e-04\\
		
		& & & H2 3.6$\mu$m & 3.39e-05 & H2 4.41$\mu$m & 1.52e-04 & H2 8.02$\mu$m & 3.61e-04  & H2 1.95$\mu$m & 4.31e-04\\  
		
		& & & [NI] 5200$\AA$ & 3.34e-05 & H2 8.0$\mu$m & 1.50e-04 & H2 6.9$\mu$m & 3.22e-04 & CO 260$\mu$m & 4.26e-04\\  
		\hline
	 	\hline
	\end{tabular}}
\\
\end{table*}	
\begin{table*}[htbp!]
\centering
     \caption{Brightest predicted {\bf cumulative} line emission.}
     \setlength{\tabcolsep}{4pt}
       {\small \begin{tabular}{ l l l | l c | l c | l c | l c}
       		\hline
       		\hline
        		 \noalign{\smallskip}
       		 \multicolumn{3}{l |}{Model} & \multicolumn{2}{c |}{\av = 0.1} & \multicolumn{2}{c |}{\av = 1} & \multicolumn{2}{c |}{\av = 5} & \multicolumn{2}{c}{\av = 30}\\
		\noalign{\smallskip}
			\hline
			 \gx & v$_{\rm tur}$ & Z & line & emission & line & emission & line & emission & line & emission\\
			  & km/s & Z$_{\odot}$ & & erg s$^{-1}$ cm$^{-2}$ & & erg s$^{-1}$ cm$^{-2}$ & & erg s$^{-1}$ cm$^{-2}$ & & erg s$^{-1}$ cm$^{-2}$\\
			\hline
			\noalign{\smallskip}
		10 & 50 & 1 & [OII] 3729$\AA$ & 1.69e-04  & \cii & 7.28e-04 & \cii & 1.26e-03 & \cii & 1.42e-03\\ 
		
		& & & \cii & 1.35e-04 & [OI] 63$\mu$m &  2.94e-04 & [OI] 63$\mu$m & 8.05e-04 & [OI] 63$\mu$m & 1.16e-03\\
		
		& & & [NII] 6583$\AA$ & 1.24e-04 & [OII] 3729$\AA$ & 1.44e-04 & [SiII] 35$\mu$m & 2.85e-04 & [CI] 370$\mu$m & 5.52e-04\\  
		
		& & & [OII] 3726$\AA$ & 1.14e-04 & [SIII] 33$\mu$m  &  1.35e-04 & [NeII] 13$\mu$m & 1.65e-04 & [CI] 609$\mu$m & 4.34e-04\\  
		
		& & & [OI] 6300$\AA$ & 8.41e-05 & [SiII] 35$\mu$m &  1.34e-04 & [CI] 370$\mu$m & 1.59e-04 & [SiII] 35$\mu$m & 2.79e-04\\  
		
		& & & HI 6563$\AA$ & 7.94e-05  & [NII] 6583$\AA$ &  1.23e-04 & [OII] 3729$\AA$  & 1.34e-04 & H2 6.9$\mu$m & 1.50e-04\\
		
		& & & [SII] 6716$\AA$ & 7.53e-05  & [OI] 6300$\AA$ & 1.21e-04 & [SIII] 33$\mu$m & 1.31e-04 & H2 17$\mu$m & 1.47e-04\\ 
		
		& & & [SIII] 9531$\AA$ & 5.69e-05 & [SII] 6716$\AA$  & 1.13e-04 & H2 17$\mu$m & 1.26e-04 & [OII] 3729$\AA$ & 1.34e-04\\  
		
		& & & [SIII] 33$\mu$m & 5.33e-05  & [NeII] 13$\mu$m & 1.11e-04 & HI 6563$\AA$ & 1.11e-04 & [NeII] 13$\mu$m & 1.22e-04\\  
		
		& & & [SII] 6731$\AA$ & 5.22e-05 & HI 6563$\AA$ &  1.07e-04 & [NII] 6583$\AA$ & 1.09e-04 & H2 5.5$\mu$m & 1.11e-04\\  
		
		& & & [OI]  63$\mu$m & 4.26e-05 & [FeII] 5.3$\mu$m &  9.93e-05 & [CI] 609$\mu$m & 1.09e-04 & [NII] 6583$\AA$ & 1.08e-04\\ 
		
		& & & [NII] 6548$\AA$ & 4.22e-05 & [OII] 3726$\AA$ & 9.70e-05 & H2 9.66$\mu$m  &  1.07e-04 & [SIII] 33$\mu$m & 1.07e-04\\
		
		& & & HeII 1640$\AA$ & 3.83e-05 & [SIII] 9531$\AA$ & 8.73e-05 & [OI] 6300$\AA$ & 1.05e-04 & [OI] 6300$\AA$ & 1.04e-04\\  
		
		& & & [NeIII] 15$\mu$m & 3.38e-05 & [SII] 6731$\AA$ & 7.80e-05 & [FeII] 5.3$\mu$m & 1.00e-04 & HI 6563$\AA$ & 1.04e-04\\  
		
		& & & [FeII] 5.3$\mu$m &  3.17e-05 & [NeIII] 15$\mu$m & 7.34e-05 & [SII] 6716$\AA$ & 9.70e-05 & H2 9.66$\mu$m & 1.03e-04\\  
		
		& & & [SiII] 35$\mu$m &  3.05e-05 & [SIII] 19$\mu$m & 7.11e-05 & [OII] 3726$\AA$ & 9.08e-05 & [OI] 145$\mu$m & 9.73e-05\\ 
		
		& & & [SIII] 19$\mu$m &  3.01e-05 & H2 9.66$\mu$m & 6.22e-05 & [NeIII] 15$\mu$m & 8.03e-05 & [SII] 6716$\AA$ & 9.59e-05\\ 
		
		& & & HeII 304$\AA$ &   2.98e-05 & [FeII] 1.2$\mu$m & 5.58e-05 & [SIII] 9531$\AA$ & 7.40e-05 & [OII] 3726$\AA$ & 9.06e-05\\
 		
 		& & & [NI] 5200$\AA$ &  2.98e-05 & [FeII] 1.6$\mu$m & 4.95e-05 & [OI] 145$\mu$m & 7.24e-05 & [FeII] 5.3$\mu$m & 7.73e-05\\  
		
		& & & [OI] 6364$\AA$ & 2.69e-05 & [ArII] 7.0$\mu$m & 4.71e-05 & [FeII] 26$\mu$m  & 7.06e-05 & H2 4.69$\mu$m & 7.50e-05\\  
		\noalign{\smallskip}
		\hline
		\noalign{\smallskip}
		10 & 10 & 0.3 & \cii & 3.71e-04 & \cii & 1.06e-03 & H2 6.9$\mu$m & 1.58e-03  &  H2 6.9$\mu$m & 4.97e-03\\  
		
		& & & [OI] 63$\mu$m & 2.65e-04 & [OI] 63$\mu$m & 7.46e-04 & H2 5.5$\mu$m & 1.53e-03 & H2 5.5$\mu$m & 4.51e-03\\   
		
		& & & HI 6563$\AA$ & 1.23e-04 & H2 6.9$\mu$m & 2.35e-04 & [OI] 63$\mu$m & 1.28e-03 & H2 4.7$\mu$m & 3.41e-03\\  
		
		& & & [SII] 6716$\AA$ & 7.28e-05 & H2 5.5$\mu$m & 2.27e-04  & \cii & 1.26e-03 & H2 3.83$\mu$m & 2.28e-03\\   
		
		& & & [SiII] 35$\mu$m & 7.01e-05 & H2 9.66$\mu$m & 2.06e-04 & H2 4.7$\mu$m  & 1.23e-03  & H2 9.66$\mu$m & 1.69e-03\\  
		
		& & & [OI] 6300$\AA$ & 6.08e-05 & H2 4.7$\mu$m &  1.85e-04 & H2 9.66$\mu$m & 9.79e-04 & H2 17$\mu$m & 1.69e-03\\ 
		
		& & & [CI] 9850$\AA$ & 5.74e-05 & H2 17$\mu$m & 1.74e-04  & H2 4.17$\mu$m & 8.75e-04 & [OI] 63$\mu$m & 1.68e-03\\  
		
		& & & H2 9.66$\mu$m & 5.69e-05 & HI 6563$\AA$ & 1.70e-04  & H2 3.83$\mu$m & 6.93e-04 & H2 4.4$\mu$m & 1.66e-03\\   
		
		& & & H2 17$\mu$m & 5.27e-05 & H2 4.2$\mu$m & 1.35e-04 & H2 4.06$\mu$m & 6.70e-04 & [CI] 370$\mu$m & 1.65e-03\\   
		
		& & & [SII] 6731$\AA$ & 5.13e-05 & [SiII] 35$\mu$m &  1.21e-04  & H2 17$\mu$m & 6.12e-04  & H2 3.84$\mu$m & 1.59e-03\\  
		
		& & & [NeII] 13$\mu$m & 4.30e-05 & [CI] 370$\mu$m &  1.18-04 & H2 3.84$\mu$m & 6.07e-04 & \cii & 1.57e-03\\  
		
		& & & HI 4861$\AA$ & 3.96e-05 & H2 3.83$\mu$m  &  1.17e-04 & H2 4.4$\mu$m & 6.01e-04 & H2 6.1$\mu$m & 1.51e-03\\ 
		
		& & & [OII] 3729$\AA$ & 3.02e-05 & H2 4.06$\mu$m & 1.06e-04 & [CI] 370$\mu$m & 5.49e-04 & H2 4.3$\mu$m & 1.46e-03\\  
		
		& & & [NII] 6583$\AA$  & 2.90e-05  & H2 1.7$\mu$m &  9.98e-05 & H2 1.7$\mu$m & 5.07e-04 & H2 4.9$\mu$m & 1.27e-03\\  
		
		& & & H2 12$\mu$m & 2.86e-05 & H2 3.84$\mu$m & 9.70e-05 & H2 4.3$\mu$m & 4.54e-04 & H2 5.0$\mu$m & 1.21e-03\\ 
		
		& & & [FeII] 5.3$\mu$m & 2.80e-05 & H2 4.4$\mu$m &  9.18e-05 & H2 3.6$\mu$m & 4.40e-04 & H2 3.6$\mu$m & 1.14e-03\\ 
		
		& & & [NI] 5200$\AA$ & 2.66e-05 & H2 3.6$\mu$m  & 7.34e-05 & H2 4.9$\mu$m & 4.04e-04 & H2 8.0$\mu$m & 1.10e-03\\ 
		
		& & & [OI] 145$\mu$m &  2.23e-05 & H2 4.3$\mu$m & 7.05e-05 & H2 6.1$\mu$m & 3.71e-04 & H2 4.4$\mu$m & 1.10e-03\\
		
		& & & [FeII] 1.2$\mu$m & 2.12e-05 & H2 1.12$\mu$m & 6.89e-05 & H2 3.48$\mu$m & 3.47e-04 & H2 3.48$\mu$m & 9.31e-04\\  
		
		& & & [OII] 3726$\AA$ & 2.12 e-05 & [SII] 6716$\AA$ & 6.07e-05 & H2 2.5$\mu$m & 3.45e-04 & H2 3.4$\mu$m & 8.57e-04\\ 
		\hline
	 	\hline
	\end{tabular}}
\\
\end{table*}
\begin{table*}[htbp!]
\centering
     \caption{Brightest predicted {\bf cumulative} line emission.}
     \setlength{\tabcolsep}{4pt}
       {\small \begin{tabular}{ l l l | l c | l c | l c | l c}
       		\hline
       		\hline
        		 \noalign{\smallskip}
       		 \multicolumn{3}{l |}{Model} & \multicolumn{2}{c |}{\av = 0.1} & \multicolumn{2}{c |}{\av = 1} & \multicolumn{2}{c |}{\av = 5} & \multicolumn{2}{c}{\av = 30}\\
		\noalign{\smallskip}
			\hline
			 \gx & v$_{tur}$ & Z & line & emission & line & emission & line & emission & line & emission\\
			  & km/s & Z$_{\odot}$ & & erg s$^{-1}$ cm$^{-2}$ & & erg s$^{-1}$ cm$^{-2}$ & & erg s$^{-1}$ cm$^{-2}$ & & erg s$^{-1}$ cm$^{-2}$\\
			\hline
			\noalign{\smallskip}
		10 & 10 & 0.65 & \cii & 3.81e-04  & \cii & 1.01e-03  & [OI]  63$\mu$m & 1.28e-03 & H2 6.9$\mu$m & 2.20e-03\\     
		
		& & & [OI]  63$\mu$m & 2.82e-04  & [OI] 63$\mu$m &  7.98e-04 & \cii & 1.16e-03  & H2 5.5$\mu$m & 1.98e-03\\    
		
		& & & [SII] 6716$\AA$ &  9.84e-05 & [SiII] 35$\mu$m & 1.60e-04 & [CI] 370$\mu$m & 5.62e-04 & [OI]  63$\mu$m & 1.55e-03\\    
		
		& & & HI 6563$\AA$ &  9.82e-05 & HI 6563$\AA$ &  1.41e-04 & H2 6.9$\mu$m & 5.14e-04  & H2 4.69$\mu$m & 1.49e-03\\    
		
		& & & [SiII] 35$\mu$m & 9.31e-05 & [CI] 370$\mu$m & 1.32e-04 & H2 5.5$\mu$m  & 4.89e-04  & [CI] 370$\mu$m  & 1.39e-03\\   
		
		& & & [CI] 9850$\AA$ & 9.08e-05  & [SII] 6716$\AA$ & 8.22e-05 & H2 4.7$\mu$m & 3.86e-04 & \cii & 1.22e-03\\ 
		
		& & & [OI] 6300$\AA$ & 7.82e-05 & [CI] 9850$\AA$ & 7.83e-05 & H2 9.66$\mu$m & 3.19e-04 &  H2 4.17$\mu$m & 9.88e-04\\   
		
		& & & [SII] 6731$\AA$ & 6.70e-05 & H2 6.9$\mu$m & 6.70e-05 & H2 4.2$\mu$m & 2.74e-04 & H2 9.66$\mu$m & 7.28e-04\\   
		
		& & & [NeII] 13$\mu$m & 5.41e-05  & [OI] 6363$\AA$ & 6.50e-05  & [CI] 609$\mu$m &  2.52e-04 & H2 4.06$\mu$m & 7.18e-04\\  
		
		& & & [NII] 6583$\AA$ & 5.06e-05 & H2 9.66$\mu$m & 6.19e-05 & H2 3.83$\mu$m & 2.04e-04  & H2 3.83$\mu$m & 7.09e-04\\   
		
		& & & [OII] 3729$\AA$ & 5.05e-05 & [SII] 6731$\AA$ & 5.84e-05 & H2 4.06$\mu$m & 1.99e-04  & H2 17$\mu$m & 7.05e-04\\   
		
		& & & [FeII] 5.3$\mu$m & 3.81e-05 & H2 5.5$\mu$m & 5.82e-05 & H2 3.84$\mu$m & 1.90e-45 & H2 4.4$\mu$m & 6.85e-04\\ 
  		
  		& & & [OII] 3726$\AA$ & 3.56e-05  & [OI] 145$\mu$m & 5.46e-05 & H2 4.4$\mu$m & 1.83e-04 & H2 3.84$\mu$m  & 6.55e-04\\  
		
		& & & [NI] 5200$\AA$ & 3.26e-05 & [NeII] 13$\mu$m & 5.32e-05 & [SiII] 35$\mu$m & 1.75e-04 & [CI] 609$\mu$m & 6.12e-04\\  
		
		& & & HI 4861$\AA$ & 3.19e-05  & [CI] 609$\mu$m & 5.05e-05 & H2 17$\mu$m & 1.74e-04  & H2 6.1$\mu$m & 5.94e-04\\   
		
		& & & [CI] 9824$\AA$ & 2.38e-05 & H2 4.69$\mu$m & 4.54e-05 & H2 1.74$\mu$m & 1.62e-04 & H2 4.3$\mu$m & 5.21e-04\\   
		
		& & & [FeII] 1.2$\mu$m & 2.39e-05 & HI 4861$\AA$ & 4.23e-05 & H2 3.6$\mu$m & 1.340e-04  & H2 5.05$\mu$m & 5.12e-04\\   
		
		& & & H2 9.66$\mu$m & 2.32e-05 & H2 17$\mu$m & 4.24e-05 & H2 4.3$\mu$m & 1.37e-04 & H2 4.9$\mu$m & 5.00e-04\\ 
		
		& & & [OI] 6363$\AA$ & 2.24e-05 & [NII] 6583$\AA$ & 4.23e-05  & HI 6563$\AA$ & 1.34e-04  & H2 3.6$\mu$m & 4.63e-04\\   
		
		& & & [OI] 145$\mu$m & 2.21e-05 & [OII] 3729$\AA$ & 4.09e-05 & H2 6.1$\mu$m & 1.31e-04  & H2 8.0$\mu$m & 4.38e-04\\   
		\hline
	 	\hline
	\end{tabular}}
\\
\end{table*}

\end{document}